# The single edge notch fracture test for viscoelastic elastomers

Farhad Kamarei, Fabio Sozio, and Oscar Lopez-Pamies

Department of Civil and Environmental Engineering, University of Illinois, Urbana–Champaign, USA

Making use of the Griffith criticality condition recently introduced by Shrimali and Lopez-Pamies, this work presents a comprehensive analysis of the single edge notch fracture test for viscoelastic elastomers. The results—comprised of a combination of a parametric study and direct comparisons with experiments—reveal how the non-Gaussian elasticity, the nonlinear viscosity, and the intrinsic fracture energy of elastomers interact and govern when fracture nucleates from the pre-existing crack in these tests. The results also serve to quantify the limitations of existing analyses, wherein viscous effects and the actual geometries of the pre-existing cracks and the specimens are neglected.

## 1 Introduction

As a direct consequence of two elementary observations, Shrimali and Lopez-Pamies (2023a) have shown that the original form (Rivlin and Thomas 1953; Greensmith and Thomas 1955)

$$-\frac{\partial \mathcal{W}}{\partial \Gamma_0} = T_c \tag{1}$$

of the Griffith criticality condition that describes the growth of cracks in viscoelastic elastomers[1] subjected to quasi-static mechanical loads can be reduced to the fundamental form

$$-\frac{\partial \mathcal{W}^{\mathrm{Eq}}}{\partial \Gamma_0} = G_c. \tag{2}$$

In Equation (1), the left-hand side $-\partial \mathcal{W}/\partial \Gamma_0$ denotes the change[2] in total deformation—stored and dissipated[3]—energy

$$\mathcal{W} = \underbrace{\mathcal{W}^{\mathrm{Eq}} + \mathcal{W}^{\mathrm{NEq}}}_{\text{stored}} + \underbrace{\mathcal{W}^{v}}_{\text{dissipated}} \tag{3}$$

[1] By "viscoelastic elastomers" we mean elastomers that in addition to being capable to store energy by elastic deformation, they are capable to dissipate energy by viscous deformation. Dissipation mechanisms other than viscous deformation and fracture (e.g., Mullins effect, strain-induced crystallization) are considered to be negligible or absent altogether.

[2] The derivative $-\partial \mathcal{W}/\partial \Gamma_0$ is to be carried out under conditions of fixed deformation of those parts of the boundary that are not traction-free so that no work is done by the external forces.

[3] The stored part of the energy is comprised of an equilibrium part $\mathcal{W}^{\mathrm{Eq}}$ and a non-equilibrium part $\mathcal{W}^{\mathrm{NEq}}$, the latter representing the part of the stored energy that gets dissipated via viscous deformation eventually. For example, for elastomers whose viscoelastic behavior can be described by the Zener model depicted in Figure 2, in terms of an equilibrium stored-energy function $\psi^{\mathrm{Eq}}$, a non-equilibrium stored-energy function $\psi^{\mathrm{NEq}}$, and a dissipation potential $\phi$, the three parts in the energy decomposition (3) are given by the integrals

$$\mathcal{W}^{\mathrm{Eq}} = \int_{\Omega_0} \psi^{\mathrm{Eq}}\, \mathrm{d}\mathbf{X}, \qquad \mathcal{W}^{\mathrm{NEq}} = \int_{\Omega_0} \psi^{\mathrm{NEq}}\, \mathrm{d}\mathbf{X}, \qquad \mathcal{W}^{v} = \int_{\Omega_0}\int_0^T 2\phi\, \mathrm{d}\tau\, \mathrm{d}\mathbf{X}$$

over the volume $\Omega_0$ occupied by the elastomer in its undeformed configuration to be evaluated at the fields that satisfy the corresponding initial-boundary-value problem; Section 3 describes these calculations in detail for the case of the single edge notch fracture test.

in the specimen at hand with respect to an added surface area $d\Gamma_0$ to the pre-existing crack $\Gamma_0$ in its reference state, while the right-hand side $T_c$ stands for the critical tearing energy, a characteristic property of the elastomer that depends on the loading history. Traditionally, the critical tearing energy has been written in the form (Gent and Schultz 1972; Gent 1996)

$$T_c = G_c(1 + f_c), \tag{4}$$

where the material constant $G_c$ denotes the intrinsic fracture energy, or critical energy release rate, associated with the creation of new surface in the given elastomer, while $f_c$ stands for a non-negative material function of the loading history.

In expression (2), on the other hand, the left-hand side $-\partial\mathcal{W}^{\rm Eq}/\partial\Gamma_0$ denotes the change *not* in $\mathcal{W}$ but only in equilibrium elastic energy $\mathcal{W}^{\rm Eq}$ stored in the specimen with respect to an added surface area to the pre-existing crack $\Gamma_0$, while the right-hand side involves *not* $T_c$ but only the intrinsic fracture energy $G_c$ of the elastomer.

From a fundamental point of view, the criticality condition (2) states that whether an elastomer deforms or creates new surface from a pre-existing crack is dictated by a competition solely between its stored equilibrium elastic energy and its intrinsic fracture energy. A corollary of this result is that the critical tearing energy $T_c$ in (1)—whose dependence via $f_c$ on the loading history had remained unknown for decades, preventing the use of the criticality condition (1) as a predicting theory—is given by the formula

$$T_c = G_c(1 + f_c) = G_c - \frac{\partial\mathcal{W}^{\rm NEq}}{\partial\Gamma_0} - \frac{\partial\mathcal{W}^{v}}{\partial\Gamma_0}, \tag{5}$$

where the right-hand side is to be evaluated at the instance in time at which (2) is attained along the loading path of interest. This last relation clarifies how $T_c$ describes the total energy (per unit undeformed fracture area) expended in the tearing process, as it makes explicit the contributions by the creation of new surface ($G_c$) and by viscous effects ($-\partial\mathcal{W}^{\rm NEq}/\partial\Gamma_0 - \partial\mathcal{W}^{v}/\partial\Gamma_0$).

From a practical point of view, the criticality condition (2) has the merit of being straightforward to employ for arbitrary specimen geometries and arbitrary loading conditions. This is because its evaluation requires having knowledge of the finite viscoelastic behavior of the elastomer—from which the stored equilibrium elastic energy $\mathcal{W}^{\rm Eq}$, as well as the stored non-equilibrium elastic energy $\mathcal{W}^{\rm NEq}$ and the dissipated viscous energy $\mathcal{W}^{v}$, can be computed—and of its intrinsic fracture energy. Both of these are material properties that can be measured once and for all.

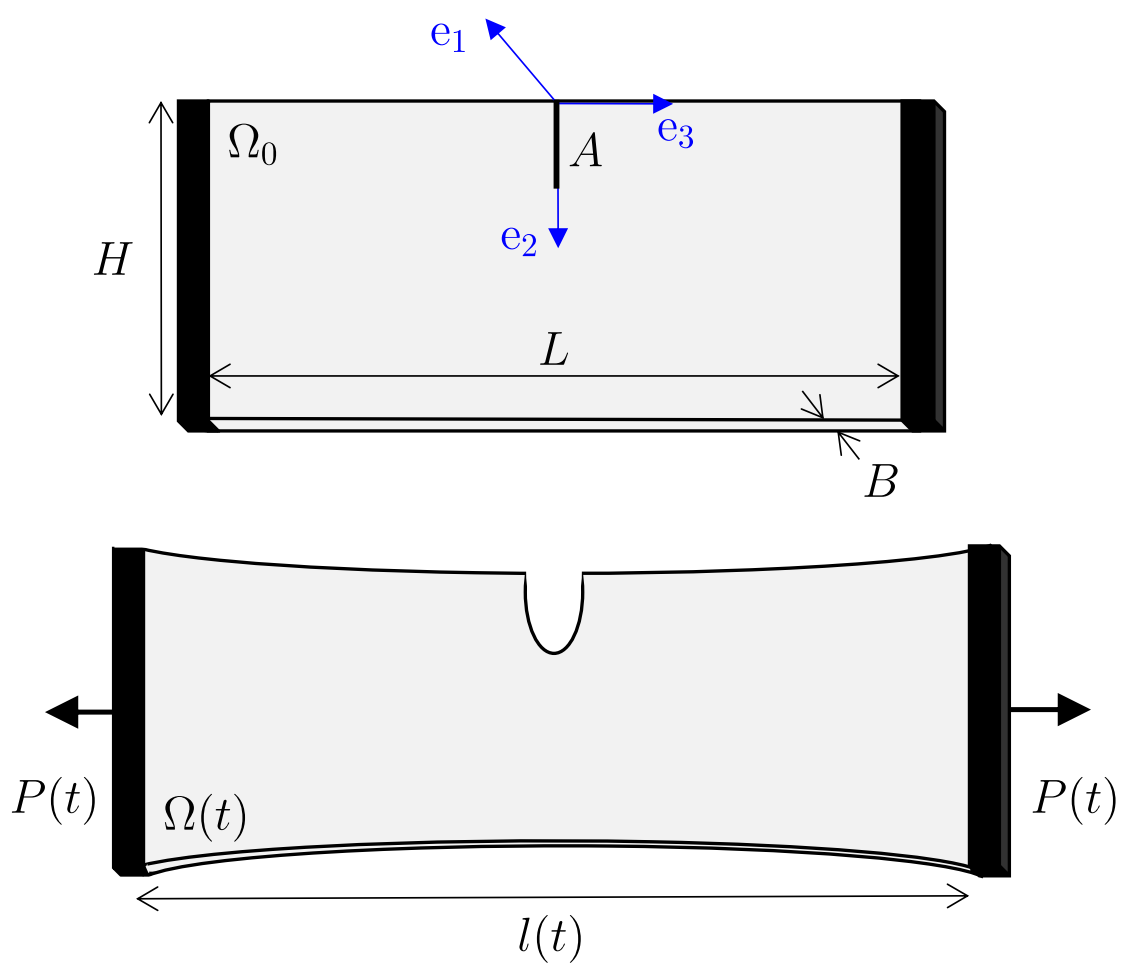

**Figure 1** Schematic of the single edge notch fracture test for a viscoelastic elastomer illustrating the geometry of the specimen, the applied time-dependent separation $l(t)$ between the grips, and the corresponding force $P(t)$ at the grips. It is standard practice to use the normalized quantities $\Lambda = l(t)/L$ and $S = P(t)/(HB)$. We refer to them, respectively, as the global stretch and the global nominal stress. The focus of this work is on initial sizes $A$ of the crack (the "notch") that are larger than the initial thickness $B$ of the specimen.

As a first application of the criticality condition (2), Shrimali and Lopez-Pamies (2023a) employed it to explain the "pure-shear" fracture test. In two ensuing companion papers (Lopez-Pamies and Shrimali 2023; Shrimali and Lopez-Pamies 2023b), the same authors made use of (2) to explain two other types of prominent tests used to study fracture in viscoelastic elastomers: the delayed fracture test and the trousers fracture test. More recently, the criticality condition (2) has been recast into a regularized phase-field formulation (Kamarei et al. 2025) and shown to explain additional experimental observations.

The objective of this paper is to make use of the criticality condition (2) to explain in a detailed and quantitative manner yet another of the staple fracture tests utilized in the literature to probe the growth of cracks in elastomers, that depicted schematically in Figure 1: *the single edge notch fracture test*.

## 1.1 A comment on previous works on the single edge notch fracture test

The first experimental results reported in the literature of single edge notch fracture tests applied to elastomers date back to the 1930s. They correspond to a series of tests carried out by Busse (1934) with specimens made of natural rubber that featured a wide range of sizes for the pre-existing crack.

The early work by Busse (1934) went unnoticed in the mechanics community until Rivlin and Thomas (1953) presented their own single edge notch fracture tests alongside a first theoretical analysis in the 1950s. Their analysis amounts to the approximate evaluation of the criticality condition (1) making use *not* of local fields but of global quantities—in particular, the separation $l(t)$ between the grips and the corresponding force $P(t)$ at the grips—and of two simplifying *constitutive* and *geometric* assumptions:

1. That the mechanical behavior of the elastomer that the specimen is made of is purely elastic; and
2. That the size $A$ of the pre-existing crack is much smaller than the length $L$ and the width $H$ of the specimen, but much larger than its thickness $B$ (see Figure 1), so that boundary effects can be neglected and plane stress conditions apply.

Under these assumptions, which, as illustrated in Section 4.4 below, are rarely satisfied in practice, they concluded that

$$-\frac{\partial \mathcal{W}}{\partial \Gamma_0} = 2AK(\Lambda)W_{\text{un}}(\Lambda) = T_c, \tag{6}$$

where $W_{\text{un}}(\Lambda)$ stands for the stored-energy function of the elastomer under uniaxial tension evaluated at the global stretch $\Lambda = l/L$, while $K(\Lambda)$ is a numerical factor that varies with $\Lambda$.

In 1963, a decade after Rivlin and Thomas (1953) introduced the result (6), Greensmith (1963) carried out single edge notch fracture tests on four different types of natural rubbers aimed at determining the numerical factor $K(\Lambda)$ in (6) directly from experimental measurements. The fitting of (6) to his data indicated that, irrespective of the type of rubber that he tested, $K(\Lambda)$ decreases nonlinearly from a value of about 3 at small stretches when $\Lambda \approx 1$ to a value somewhat below 2 at $\Lambda \approx 3$. Later, guided by the experiments of Greensmith (1963) and by his own 2D plane stress finite element (FE) simulations of the test, Lindley (1972) proposed the approximation

$$K(\Lambda) = \frac{2.95 - 0.08(\Lambda - 1)}{\sqrt{\Lambda}} \approx \frac{3}{\sqrt{\Lambda}}. \tag{7}$$

By now, experiments beyond those of Greensmith (1963) are abundant[4] in the literature (Hamed and Park 1999; Cristiano et al. 2011; Fan et al. 2017; Roucou et al. 2018; Slootman et al. 2020; Le Menn 2022; Wang and Fan 2023; Xue et al. 2024). However, most of these experiments have been carried out at a *fixed constant stretch rate*—wherein a global stretch of the form $\Lambda = 1 + \dot{\Lambda}_0 t$, with fixed $\dot{\Lambda}_0$, is prescribed—which conceals the viscous dissipation that inevitably takes place in these tests. Furthermore, many of these experiments have been carried out with specimens for which the size of the pre-existing crack is comparable to both the specimen width and the specimen thickness, that is, $A \sim H$ and $A \sim B$, so that boundary effects and the 3D character of the mechanical fields are not expected to be negligible.

Beyond the simulations of Lindley (1972), a plurality of FE simulations are also now available in the literature (Yeoh 2002; Chen et al. 2017; Kang et al. 2020; Liu et al. 2020). Invariably, all of these simulations assume that the mechanical behavior of elastomers is purely elastic. What is more, they are restricted to 2D plane stress simulations.

In this context, this work aims at reporting an analysis that fully accounts for the viscoelasticity of elastomers and the 3D geometry of the pre-existing crack and the specimen in single edge notch fracture tests, and hence one that should allow to explain and appropriately interpret the results from such tests.

[4] The single edge notch fracture test has also been of recent popular use to study fracture in elastomers containing a liquid solvent (Kwon et al. 2011; Guo et al. 2016; Lin et al. 2022).

### 1.2 Clarification of two misconceptions of the criticality condition (2)

Before proceeding with the proposed analysis of the single edge notch fracture test, it is appropriate to close this Introduction by clarifying two misconceptions regarding the criticality condition (2) that have recently appeared in the literature.

The first of these misconceptions is that the equilibrium energy $\mathcal{W}^{\text{Eq}}$ in the criticality condition (2) is *not* "*the elastic energy stored in the system at equilibrium i.e. the elastic energy after keeping the system which prevail at time t for an infinite long time with fixed boundary conditions*" as stated, for instance, in (Persson et al. 2024). Instead, $\mathcal{W}^{\text{Eq}}$ is the integral over the volume occupied by the elastomer in its undeformed configuration evaluated at the fields that satisfy the corresponding initial-boundary-value problem *at the current time t*. For example, as already noted above, for elastomers whose viscoelastic behavior is described by the Zener model depicted in Figure 2

$$\mathcal{W}^{\text{Eq}} = \int_{\Omega_0} \psi^{\text{Eq}}(I_1)\,\mathrm{d}\mathbf{X} \quad \text{with} \quad I_1 = \operatorname{tr}(\nabla\mathbf{y}^\top(\mathbf{X},t)\nabla\mathbf{y}(\mathbf{X},t)), \tag{8}$$

where the deformation field $\mathbf{y}(\mathbf{X},t)$ is solution of the equations of equilibrium at the current time $t$. By definition, the equilibrium energy $\mathcal{W}^{\text{Eq}}$ depends thus on *the entire viscoelastic behavior of the elastomer*—implicitly via the deformation field $\mathbf{y}(\mathbf{X},t)$—and not just on its behavior at equilibrium. This implies, in particular, that $\mathcal{W}^{\text{Eq}}$ and hence its derivative $-\partial\mathcal{W}^{\text{Eq}}/\partial\Gamma_0$ in the criticality condition (2) can be rate dependent.

The second misconception is that the criticality condition (2) does *not* "*only address the onset of crack propagation*" as also stated in (Persson et al. 2024). Instead, the criterion describes the growth of cracks in a *complete manner*, that is, it describes the growth of a large crack in a viscoelastic elastomer that is subjected to any loading condition, absent inertia. As such, it describes for instance the (uniform or not) speed at which a crack grows or arrests. This kind of predictions have been recently reported in (Kamarei et al. 2025), where the criterion has been recast into a regularized phase-field formulation that is amenable to computations for arbitrary loading conditions.

## 2 Formulation of the initial-boundary-value problem for the single edge notch fracture test

### 2.1 Initial configuration and kinematics

Consider the rectangular specimens depicted in Figure 1 of length $L$ and width $H$ in the $\mathbf{e}_3$ and $\mathbf{e}_2$ directions and constant thickness $B < H$ in the $\mathbf{e}_1$ direction. The specimens contain a pre-existing edge crack of length $A < H$ in the $\mathbf{e}_2$ direction. Here, $\{\mathbf{e}_i\}$ $(i = 1,2,3)$ stands for the laboratory frame of reference. We place its origin at the specimens' midplane along the edge containing the crack so that, in their initial configuration at time $t = 0$, the specimens occupy the domain

$$\overline{\Omega}_0 = \{\mathbf{X} : \mathbf{X} \in \mathcal{P}_0 \setminus \Gamma_0\}, \tag{9}$$

where

$$\mathcal{P}_0 = \{\mathbf{X} : |X_1| \leqslant B/2, 0 \leqslant X_2 \leqslant H, |X_3| \leqslant L/2\} \tag{10}$$

and

$$\Gamma_0 = \{\mathbf{X} : |X_1| \leqslant B/2, 0 \leqslant X_2 \leqslant A, X_3 = 0\}\,. \tag{11}$$

At a later time $t \in (0,T]$, in response to the boundary conditions described below, the position vector $\mathbf{X}$ of a material point occupies a new position $\mathbf{x}$ specified by an invertible mapping $\mathbf{y}$ from $\Omega_0$ to the current configuration $\Omega(t) \subset \mathbb{R}^3$. We write

$$\mathbf{x} = \mathbf{y}(\mathbf{X},t) \tag{12}$$

and the associated deformation gradient and velocity fields at $\mathbf{X} \in \Omega_0$ and $t \in (0, T]$ as

$$\mathbf{F}(\mathbf{X}, t) = \nabla \mathbf{y}(\mathbf{X}, t) = \frac{\partial \mathbf{y}}{\partial \mathbf{X}}(\mathbf{X}, t) \tag{13}$$

and

$$\mathbf{V}(\mathbf{X}, t) = \dot{\mathbf{y}}(\mathbf{X}, t) = \frac{\partial \mathbf{y}}{\partial t}(\mathbf{X}, t). \tag{14}$$

Throughout, we shall use the "dot" notation to denote the material time derivative (i.e., with $\mathbf{X}$ held fixed) of field quantities.

## 2.2 Constitutive behavior of the elastomer

The specimens are taken to be made of an isotropic incompressible elastomer. Making use of the two-potential formalism (Kumar and Lopez-Pamies 2016), we describe its constitutive behavior by two thermodynamic potentials, a free energy density, or stored-energy function, of the form

$$\psi(\mathbf{F}, \mathbf{C}^v) = \begin{cases} \psi^{\text{Eq}}(I_1) + \psi^{\text{NEq}}(I_1^e) & \text{if } J = 1 \\ +\infty & \text{else} \end{cases} \tag{15}$$

that describes how the elastomer stores energy through elastic deformation and a dissipation potential of the form

$$\phi(\mathbf{F}, \mathbf{C}^v, \dot{\mathbf{C}}^v) = \begin{cases} \frac{1}{2}\dot{\mathbf{C}}^v \cdot \mathcal{A}(\mathbf{F}, \mathbf{C}^v)\dot{\mathbf{C}}^v & \text{if } J^v = 1 \\ +\infty & \text{else} \end{cases} \tag{16}$$

with $2\mathcal{A}_{ijkl}(\mathbf{F}, \mathbf{C}^v) = \eta(\mathbf{C}, \mathbf{C}^v) C^{v\,-1}_{ik} C^{v\,-1}_{jl}$ that describes how the elastomer dissipates energy through viscous deformation. In these expressions, the symmetric second-order tensor $\mathbf{C}^v$ is an internal variable of state that stands for a measure of the "viscous part" of the deformation gradient $\mathbf{F}$,

$$I_1 = \operatorname{tr} \mathbf{C}, \quad J = \sqrt{\det \mathbf{C}}, \quad I_1^e = \operatorname{tr}(\mathbf{C}\mathbf{C}^{v-1}), \quad J^v = \sqrt{\det \mathbf{C}^v}, \tag{17}$$

where $\mathbf{C} = \mathbf{F}^\top \mathbf{F}$ denotes the right Cauchy-Green deformation tensor, and $\psi^{\text{Eq}}$, $\psi^{\text{NEq}}$, $\eta$ are any suitably well-behaved non-negative material functions of their arguments.

Granted the two thermodynamic potentials (15) and (16), it follows that the first Piola-Kirchhoff stress tensor $\mathbf{S}$ at any material point $\mathbf{X} \in \Omega_0$ and time $t \in [0, T]$ is given by the relation (Kumar and Lopez-Pamies 2016)

$$\mathbf{S}(\mathbf{X}, t) = \frac{\partial \psi}{\partial \mathbf{F}}(\mathbf{F}, \mathbf{C}^v), \tag{18}$$

where $\mathbf{C}^v$ is implicitly defined by the evolution equation

$$\begin{cases} \dfrac{\partial \psi}{\partial \mathbf{C}^v}(\mathbf{F}, \mathbf{C}^v) + \dfrac{\partial \phi}{\partial \dot{\mathbf{C}}^v}(\mathbf{F}, \mathbf{C}^v, \dot{\mathbf{C}}^v) = \mathbf{0} \\ \mathbf{C}^v(\mathbf{X}, 0) = \mathbf{I}. \end{cases} \tag{19}$$

Making use of the specific isotropic incompressible forms (15) and (16), this relation can be rewritten more explicitly as

$$\mathbf{S}(\mathbf{X}, t) = 2\psi^{\text{Eq}}_{I_1}\mathbf{F} + 2\psi^{\text{NEq}}_{I_1^e}\mathbf{F}\mathbf{C}^{v-1} - p\mathbf{F}^{-\top}, \tag{20}$$

where $p$ stands for the arbitrary hydrostatic pressure associated with the incompressibility constraint $J = 1$ of the elastomer and $\mathbf{C}^v$ is defined implicitly as the solution of the evolution equation

$$\begin{cases} \dot{\mathbf{C}}^v(\mathbf{X}, t) = \dfrac{2\psi^{\text{NEq}}_{I_1^e}}{\eta(\mathbf{C}, \mathbf{C}^v)}\left(\mathbf{C} - \frac{1}{3}(\mathbf{C} \cdot \mathbf{C}^{v-1})\mathbf{C}^v\right) \\ \mathbf{C}^v(\mathbf{X}, 0) = \mathbf{I}. \end{cases} \tag{21}$$

In these last expressions, we have made use of the notation $f_x = \mathrm{d}f(x)/\mathrm{d}x$ for the derivative of scalar-valued functions of one variable.

For a detailed account of the constitutive relation (20)-(21) and its physical interpretation, the interested reader is referred to (Kumar and Lopez-Pamies 2016). Here, it suffices to make the following remark.

**Remark 1** The constitutive relation (20)-(21) is nothing more than a generalization of the classical Zener or standard solid model (Zener 1948) to the setting of finite deformations. To see this, note that the rheological representation of the constitutive relation (20)-(21) is given by the Zener rheological representation shown in Figure 2. Note as well that in the limit of small deformations, as $\mathbf{F} \to \mathbf{I}$, the constitutive relation (20)-(21) reduces to the classical linear viscoelastic Zener model

$$\mathrm{S}(\mathbf{X}, t) = 2\mu^{\mathrm{Eq}}\boldsymbol{\varepsilon} + 2\mu^{\mathrm{NEq}}(\boldsymbol{\varepsilon} - \boldsymbol{\varepsilon}^v) - p\mathbf{I}, \tag{22}$$

where $\boldsymbol{\varepsilon} = 1/2(\mathbf{F} + \mathbf{F}^\top - 2\mathbf{I})$ stands for the infinitesimal strain tensor and $\boldsymbol{\varepsilon}^v = 1/2(\mathbf{C}^v - \mathbf{I})$ is solution of the evolution equation

$$\begin{cases} \dot{\boldsymbol{\varepsilon}}^v(\mathbf{X}, t) = \dfrac{\mu^{\mathrm{NEq}}}{\eta_0}(\boldsymbol{\varepsilon} - \boldsymbol{\varepsilon}^v - \frac{1}{3}\operatorname{tr}(\boldsymbol{\varepsilon} - \boldsymbol{\varepsilon}^v)\mathbf{I}) \\ \boldsymbol{\varepsilon}^v(\mathbf{X}, 0) = \mathbf{0}. \end{cases} \tag{23}$$

In these expressions, $\mu^{\mathrm{Eq}} = 2\psi^{\mathrm{Eq}}_{I_1}(3)$ and $\mu^{\mathrm{NEq}} = 2\psi^{\mathrm{NEq}}_{I_1^e}(3)$ stand for the initial shear moduli associated with the equilibrium and non-equilibrium stored-energy functions, respectively, while $\eta_0 = \eta(\mathbf{I}, \mathbf{I})$ stands for the initial viscosity.

**Figure 2** Rheological representation of the two-potential viscoelastic model (20)-(21).

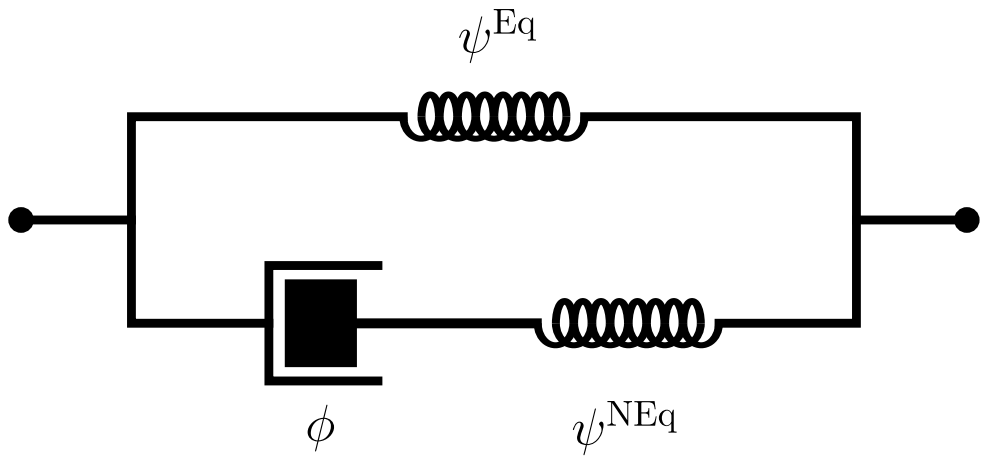

In the parametric study that is presented below, we make use of the following specific forms for the equilibrium and non-equilibrium stored-energy functions in Equation (15) and viscosity function in Equation (16):

$$\begin{cases} \psi^{\mathrm{Eq}}(I_1) = \displaystyle\sum_{r=1}^{2} \frac{3^{1-\alpha_r}}{2\alpha_r}\mu_r(I_1^{\alpha_r} - 3^{\alpha_r}) & \text{(24a)} \\ \psi^{\mathrm{NEq}}(I_1^e) = \displaystyle\sum_{r=1}^{2} \frac{3^{1-\beta_r}}{2\beta_r}\nu_r(I_1^{e\,\beta_r} - 3^{\beta_r}) & \text{(24b)} \\ \eta(\mathbf{C}, \mathbf{C}^v) = \widetilde{\eta}(I_1^v, \mathcal{J}_2^{\mathrm{NEq}}) & \text{(24c)} \end{cases}$$

with $I_1^v = \operatorname{tr}\mathbf{C}^v$ and $\mathcal{J}_2^{\mathrm{NEq}} = (I_1^{e\,2}/3 - I_2^e)(\sum_{r=1}^2 3^{1-\beta_r}\nu_r I_1^{e\,\beta_r-1})^2$, where $2I_2^e = I_1^{e\,2} - \operatorname{tr}(\mathbf{C}\mathbf{C}^{v-1}\mathbf{C}\mathbf{C}^{v-1})$, which result in the constitutive relation

$$\mathrm{S}(\mathbf{X}, t) = \sum_{r=1}^{2} 3^{1-\alpha_r}\mu_r I_1^{\alpha_r-1}\mathbf{F} + \sum_{r=1}^{2} 3^{1-\beta_r}\nu_r I_1^{e\,\beta_r-1}\mathbf{F}\mathbf{C}^{v-1} - p\mathbf{F}^{-\top} \tag{25}$$

with evolution equation

$$\begin{cases} \dot{\mathbf{C}}^v(\mathbf{X}, t) = \dfrac{\sum_{r=1}^N 3^{1-\beta_r}\nu_r I_1^{e\,\beta_r-1}}{\eta(\mathbf{C}, \mathbf{C}^v)}(\mathbf{C} - \frac{1}{3}(\mathbf{C}\cdot\mathbf{C}^{v-1})\mathbf{C}^v) \\ \mathbf{C}^v(\mathbf{X}, 0) = \mathbf{I}. \end{cases} \tag{26}$$

The constitutive prescription (25)-(26) includes several fundamental constitutive relations as special cases. In particular, it includes the cases of:

- A Neo-Hookean solid ($\mu_2 = \nu_1 = \nu_2 = 0$, $\alpha_1 = 1$, $\eta = 0$);
- A Newtonian fluid ($\mu_1 = \mu_2 = \nu_2 = 0$, $\nu_1 = +\infty$, $\eta = \eta_0$);
- As well as of a viscoelastic elastomer with Gaussian elasticity and constant viscosity ($\mu_2 = \nu_2 = 0$, $\alpha_1 = \beta_1 = 1$, $\eta = \eta_0$).

The prescription (25)-(26) has the additional merit of being able to describe the viscoelastic behavior of a wide range of elastomers, which typically exhibit non-Gaussian elasticity (Treloar 2005; Lopez-Pamies 2010), as well as a strongly nonlinear viscosity that depends both on the deformation and on the deformation rate (Doi and Edwards 1998; Bergstrom 1998; Amin et al. 2006; Kumar and Lopez-Pamies 2016; Zhou et al. 2018; Ghosh et al. 2021; Chockalingam et al. 2021; Chen and Ravi-Chandar 2022; Ricker et al. 2023). In particular, the viscosity of elastomers typically increases with increasing deformation and decreases with increasing deformation rate, the latter behavior being commonly referred to as shear-thinning. The viscosity function $\eta(\mathbf{C}, \mathbf{C}^v) = \widetilde{\eta}(I_1^v, \mathcal{J}_2^{\rm NEq})$ in (24) can be suitably prescribed to describe both of these behaviors via its dependence on deformation through $I_1^v$ and on deformation rate through $\mathcal{J}_2^{\rm NEq}$. In the parametric study that is presented next, we will use the specific constitutive prescription (Kumar and Lopez-Pamies 2016; Ghosh et al. 2021)

$$\widetilde{\eta}(I_1^v, \mathcal{J}_2^{\rm NEq}) = \widetilde{\eta}_\infty(I_1^v) + \frac{\widetilde{\eta}_0(I_1^v) - \widetilde{\eta}_\infty(I_1^v)}{1 + (K_2 \mathcal{J}_2^{\rm NEq})^{\gamma_2}} \tag{27}$$

with $\widetilde{\eta}_0(I_1^v) = \eta_0 + (N_0 - \eta_0)[\tanh(I_1^{v\gamma_1} - 3^{\gamma_1} - K_1) + \tanh K_1]/(1 + \tanh K_1)$ and $\widetilde{\eta}_\infty(I_1^v) = \eta_\infty$. In total, the constitutive prescription (25)-(26) with (27) contains 15 material constants. Four of them, $\mu_r$ and $\alpha_r$ ($r = 1, 2$), serve to characterize the non-Gaussian elasticity of the elastomer at states of thermodynamic equilibrium. Another four, $\nu_r$ and $\beta_r$ ($r = 1, 2$), characterize the non-Gaussian elasticity at non-equilibrium states. Finally, the last seven constants, $\eta_0$, $N_0$, $\eta_\infty$, $K_1$, $\gamma_1$, $K_2$, $\gamma_2$, serve to characterize the nonlinear deformation-dependent shear-thinning viscosity.

## 2.3 Initial and boundary conditions

In the initial configuration $\Omega_0$, the elastomer is presumed to be undeformed and stress free. This implies the initial conditions

$$\begin{cases} \mathbf{y}(\mathbf{X}, 0) = \mathbf{X} \\ p(\mathbf{X}, 0) = \mu^{\rm Eq} + \mu^{\rm NEq}, \qquad \mathbf{X} \in \overline{\Omega}_0, \\ \mathbf{C}^v(\mathbf{X}, 0) = \mathbf{I} \end{cases} \tag{28}$$

for the deformation field $\mathbf{y}(\mathbf{X}, t)$, the pressure field $p(\mathbf{X}, t)$, and the internal variable $\mathbf{C}^v(\mathbf{X}, t)$.

The bottom boundary

$$\partial\Omega_0^{\mathcal{B}} = \{\mathbf{X} : |X_1| \leqslant B/2, 0 \leqslant X_2 \leqslant H, X_3 = -L/2\} \tag{29}$$

and the top boundary

$$\partial\Omega_0^{\mathcal{T}} = \{\mathbf{X} : |X_1| \leqslant B/2, 0 \leqslant X_2 \leqslant H, X_3 = L/2\} \tag{30}$$

of the specimens are held firmly by stiff grips that are pulled apart in the $\pm\mathbf{e}_3$ directions resulting in a separation $l(t)$ between the grips. Consistent with the manner in which the vast majority of experiments are carried out, we restrict attention to the case when the grips are pulled apart at a constant rate $\dot{l}_0$, that is, when the separation between the grips is given by $l(t) = L + \dot{l}_0 t$. This results in tests carried out at the constant global stretch rate $\dot{\Lambda}_0 = \dot{l}_0/L$. Therefore, making use of the notation $\partial\Omega_0^{\mathcal{N}} = \partial\Omega_0 \setminus (\partial\Omega_0^{\mathcal{B}} \cup \partial\Omega_0^{\mathcal{T}})$ and $\mathbf{s}(\mathbf{X}, t) = \mathbf{SN}$, where $\mathbf{N}$ stands for the outward unit normal to the boundary $\partial\Omega_0$, we have the boundary conditions

$$\begin{cases} \mathbf{y}(\mathbf{X}, t) = \overline{\mathbf{y}}^{\mathcal{B}}(\mathbf{X}, t), & (\mathbf{X}, t) \in \partial\Omega_0^{\mathcal{B}} \times [0, T] \\ \mathbf{y}(\mathbf{X}, t) = \overline{\mathbf{y}}^{\mathcal{T}}(\mathbf{X}, t), & (\mathbf{X}, t) \in \partial\Omega_0^{\mathcal{T}} \times [0, T] \\ \mathbf{s}(\mathbf{X}, t) = \mathbf{0}, & (\mathbf{X}, t) \in \partial\Omega_0^{\mathcal{N}} \times [0, T] \end{cases} \tag{31}$$

with

$$\begin{cases} \overline{y}_1^{\mathcal{B}}(\mathbf{X},t)=X_1 \\ \overline{y}_2^{\mathcal{B}}(\mathbf{X},t)=X_2 \\ \overline{y}_3^{\mathcal{B}}(\mathbf{X},t)=X_3-\frac{\dot{l}_0}{2}t \end{cases} \quad \text{and} \quad \begin{cases} \overline{y}_1^{\mathcal{T}}(\mathbf{X},t)=X_1 \\ \overline{y}_2^{\mathcal{T}}(\mathbf{X},t)=X_2 \\ \overline{y}_3^{\mathcal{T}}(\mathbf{X},t)=X_3+\frac{\dot{l}_0}{2}t. \end{cases} \tag{32}$$

## 2.4 Governing equations

Neglecting inertia and body forces, the equations of balance of linear and angular momenta read as $\mathrm{Div}\,\mathbf{S}=\mathbf{0}$ and $\mathbf{S}\mathbf{F}^\top=\mathbf{F}\mathbf{S}^\top$ for $(\mathbf{X},t)\in\Omega_0\times[0,T]$. The latter is automatically satisfied (Kumar and Lopez-Pamies 2016; Sadik and Yavari 2024) by virtue of the objectivity of the thermodynamic potentials (15) and (16) and so, upon combining all the above ingredients together and introducing the notation

$$\mathbf{S}(\mathbf{X},t)=2\psi_{I_1}^{\rm Eq}\nabla\mathbf{y}+2\psi_{I_1^e}^{\rm NEq}\nabla\mathbf{y}\mathbf{C}^{v-1}-p\nabla\mathbf{y}^{-\top} \tag{33}$$

and

$$\tau(\mathbf{X},t)=\frac{\widetilde{\eta}(I_1^v,\mathcal{J}_2^{\rm NEq})}{2\psi_{I_1^e}^{\rm NEq}} \tag{34}$$

for the stress and the relaxation time function, the governing equations that describe the mechanical response of the specimens amount to the initial-boundary-value problem

$$\begin{cases} \mathrm{Div}\,\mathbf{S}=\mathbf{0}, & (\mathbf{X},t)\in\Omega_0\times[0,T] \\ \det\nabla\mathbf{y}=1, & (\mathbf{X},t)\in\Omega_0\times[0,T] \\ \mathbf{y}(\mathbf{X},t)=\overline{\mathbf{y}}^{\mathcal{B}}(\mathbf{X},t), & (\mathbf{X},t)\in\partial\Omega_0^{\mathcal{B}}\times[0,T] \\ \mathbf{y}(\mathbf{X},t)=\overline{\mathbf{y}}^{\mathcal{T}}(\mathbf{X},t), & (\mathbf{X},t)\in\partial\Omega_0^{\mathcal{T}}\times[0,T] \\ \mathbf{S}\mathbf{N}=\mathbf{0}, & (\mathbf{X},t)\in\partial\Omega_0^{\mathcal{N}}\times[0,T] \\ \mathbf{y}(\mathbf{X},0)=\mathbf{X}, & \mathbf{X}\in\Omega_0 \\ p(\mathbf{X},0)=\mu^{\rm Eq}+\mu^{\rm NEq}, & \mathbf{X}\in\Omega_0 \end{cases} \tag{35}$$

coupled with

$$\begin{cases} \dot{\mathbf{C}}^v(\mathbf{X},t)=\dfrac{1}{\tau}(\nabla\mathbf{y}^\top\nabla\mathbf{y}-\frac{1}{3}(\nabla\mathbf{y}^\top\nabla\mathbf{y}\cdot\mathbf{C}^{v-1})\mathbf{C}^v), & (\mathbf{X},t)\in\Omega_0\times[0,T] \\ \mathbf{C}^v(\mathbf{X},0)=\mathbf{I}, & \mathbf{X}\in\Omega_0 \end{cases} \tag{36}$$

for the deformation field $\mathbf{y}(\mathbf{X},t)$, the pressure field $p(\mathbf{X},t)$, and the internal variable $\mathbf{C}^v(\mathbf{X},t)$.

All the results that we present below are generated by solving numerically the initial-boundary-value problem (35)-(36) via a variant of the scheme introduced by Ghosh et al. (2021), which is based on a Crouzeix-Raviart FE discretization of space and a high-order explicit Runge-Kutta finite difference discretization of time, as well as by a related scheme that has been recently implemented as an Abaqus UMAT subroutine (Lefèvre et al. 2024); we also generated corresponding 2D plane-stress results, which turned out to be in agreement with the fully 3D results. FEniCSx and Abaqus implementations of these numerical schemes are available on GitHub.[4] In all the simulations, we made use of FE meshes that were refined around the crack front with a minimum mesh size of $h=B/10$ and loading steps that were small enough to resolve the pertinent relaxation time of the elastomer; see, e.g., Remark 10 in (Lefèvre et al. 2024).

# 3 The Griffith criticality condition (2) for the single edge notch fracture test

Having established the governing equations (35)-(36) that describe the pointwise mechanical fields in the single edge notch fracture test for viscoelastic elastomers, the evaluation of the

[4] http://pamies.cee.illinois.edu/repositories/

Griffith criticality condition (2) to determine when the pre-existing crack grows in such a test is straightforward. The pertinent calculations amount to the following three successive steps:

1. Solving the equations (35)-(36) for the deformation field $\mathbf{y}(\mathbf{X},t)$, the pressure field $p(\mathbf{X},t)$, and the internal variable $\mathbf{C}^v(\mathbf{X},t)$ for the material constants ($\mu_r$, $\alpha_r$, $\nu_r$, $\beta_r$ $\eta_0$, $N_0$, $\eta_\infty$, $K_1$, $\gamma_1$, $K_2$, $\gamma_2$) of choice, the specimen geometry $(L,H,B)$ of choice, and sets of global stretch rates and sizes of the pre-existing crack that contain the specific values $\dot{\Lambda}_0$ and $A$ of interest;
2. Computing the resulting deformation gradient $\mathbf{F}(\mathbf{X},t) = \nabla\mathbf{y}$ and associated first invariant $I_1(\mathbf{X},t) = \operatorname{tr}\mathbf{F}^T\mathbf{F}$, evaluating the equilibrium energy

$$\mathcal{W}^{\rm Eq}(\dot{\Lambda}_0,A;\Lambda) = \int_{\Omega_0} \psi^{\rm Eq}(I_1)\,{\rm d}\mathbf{X}, \tag{37}$$

and viewing the result as a function of the global stretch rate $\dot{\Lambda}_0$ and the size $A$ of the pre-existing crack, parameterized by the global stretch $\Lambda$; and

3. Computing the derivative

$$-\frac{\partial\mathcal{W}^{\rm Eq}}{\partial\Gamma_0} = -\frac{1}{B}\frac{\partial\mathcal{W}^{\rm Eq}}{\partial A}(\dot{\Lambda}_0,A;\Lambda) \tag{38}$$

of the equilibrium energy so as to finally determine from the Griffith criticality condition

$$-\frac{1}{B}\frac{\partial\mathcal{W}^{\rm Eq}}{\partial A}(\dot{\Lambda}_0,A;\Lambda_c) = G_c \tag{39}$$

the critical value $\Lambda_c$ of the global stretch $\Lambda$ at which the crack grows for the global stretch rate $\dot{\Lambda}_0$, initial crack size $A$, and critical energy release rate $G_c$ of interest.

**Remark 2** Here, it is important to emphasize that the equilibrium energy (37)—and hence its derivative (38)—depends on the *entire* viscoelastic behavior of the elastomer through the solution of the governing equations (35)-(36) for the deformation field $\mathbf{y}(\mathbf{X},t)$. Accordingly, the Griffith criticality condition (39) is expected to depend (possibly strongly) on the loading rate $\dot{\Lambda}_0$.

**Remark 3** In practice, it suffices to make use of a finite-difference approximation to compute the derivative in (38). In this work, we make use of the central three-point finite-difference quotient

$$\frac{\partial\mathcal{W}^{\rm Eq}}{\partial A} = \frac{\mathcal{W}^{\rm Eq}(\dot{\Lambda}_0,A+\Delta A;\Lambda) - \mathcal{W}^{\rm Eq}(\dot{\Lambda}_0,A-\Delta A;\Lambda)}{2\Delta A} \tag{40}$$

with $\Delta A = A/20$, which was checked to generate converged results.

Note that once the global stretch $\Lambda_c$ at which the crack grows for a given test has been determined from the Griffith criticality condition (39), it is a simple matter to determine the associated tearing energy (5). The result can be written as

$$T_c = G_c - \frac{1}{B}\frac{\partial\mathcal{W}^{\rm NEq}}{\partial A}(\dot{\Lambda}_0,A;\Lambda_c) - \frac{1}{B}\frac{\partial\mathcal{W}^{v}}{\partial A}(\dot{\Lambda}_0,A;\Lambda_c) \tag{41}$$

with

$$\mathcal{W}^{\rm NEq}(\dot{\Lambda}_0,A;\Lambda) = \int_{\Omega_0} \psi^{\rm NEq}(I_1^e)\,{\rm d}\mathbf{X} \tag{42}$$

and

$$\mathcal{W}^{v}(\dot{\Lambda}_0,A;\Lambda) = \int_{\Omega_0}\int_0^T 2\phi(\mathbf{F}(\mathbf{X},\tau),\mathbf{C}^v(\mathbf{X},\tau),\dot{\mathbf{C}}^v(\mathbf{X},\tau))\,{\rm d}\tau\,{\rm d}\mathbf{X}, \tag{43}$$

where the integrals in (42) and (43) are to be computed—much like that in (37)—in terms of the numerical solution of equations (35)-(36). Alternatively,

$$T_c = -\frac{1}{B}\frac{\partial\mathcal{W}}{\partial A}(\dot{\Lambda}_0,A;\Lambda_c), \tag{44}$$

where

$$\mathcal{W}(\dot{\Lambda}_0, A; \Lambda) = \int_L^{\Lambda L} P \, \mathrm{d}l \tag{45}$$

stands for the total deformation (stored and dissipated) energy, written here in terms of the area under the force-deformation ($P$ vs. $l$) response of the specimen.

By way of an example, Figure 3 provides a representative result for the equilibrium stored-energy function $\psi^{\mathrm{Eq}}(I_1)$ over the initial $\Omega_0$ and the current $\Omega(t)$ configurations of a specimen, with pre-existing crack of length $A = 2$ mm, at a global stretch $\Lambda = 1.5$, that has been stretched at the global stretch rate $\dot{\Lambda}_0 = 10^0$ s$^{-1}$; see also Figure A.1. Figure 4 presents corresponding results for the equilibrium energy (37) and its derivative (38) plotted as functions of the global stretch rate $\dot{\Lambda}_0$ and the initial crack size $A$ at a fixed global stretch $\Lambda$. Finally, Figure 5 presents the derivative (38) of the equilibrium energy as a function of the global stretch $\Lambda$ and the global stretch rate $\dot{\Lambda}_0$ for a fixed initial crack size $A$. The results in Figures 3 to 5 pertain to one of the single edge notch fracture tests considered in Section 4, that of an elastomer with material constants $\mu_1 = 0.09$ MPa, $\alpha_1 = 0.5$, $\mu_2 = 0.01$ MPa, $\alpha_2 = 2.5$, $\nu_1 = 1.8$ MPa, $\beta_1 = 0.5$, $\nu_2 = 0.2$ MPa, $\beta_2 = 7.5$, $\eta_0 = N_0 = \eta_\infty = 2$ MPa s, and specimens of length $L = 15$ mm, width $H = 4$ mm, and thickness $B = 0.5$ mm.

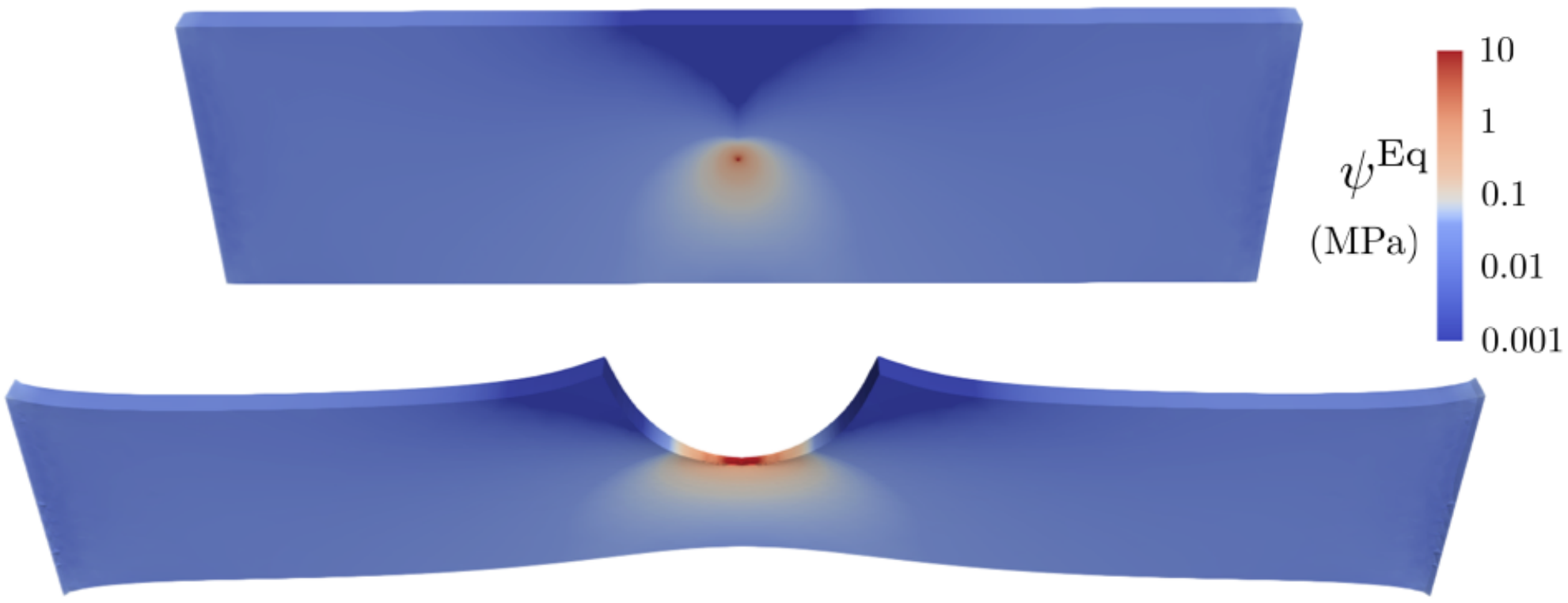

**Figure 3** Representative FE solution for the equilibrium stored-energy function $\psi^{\mathrm{Eq}}(I_1)$ (shown in MPa) over the initial $\Omega_0$ and current $\Omega(t)$ configurations of a specimen, with pre-existing crack of length $A = 2$ mm, at a global stretch $\Lambda = 1.5$, that has been stretched at the global stretch rate $\dot{\Lambda}_0 = 1$ s$^{-1}$.

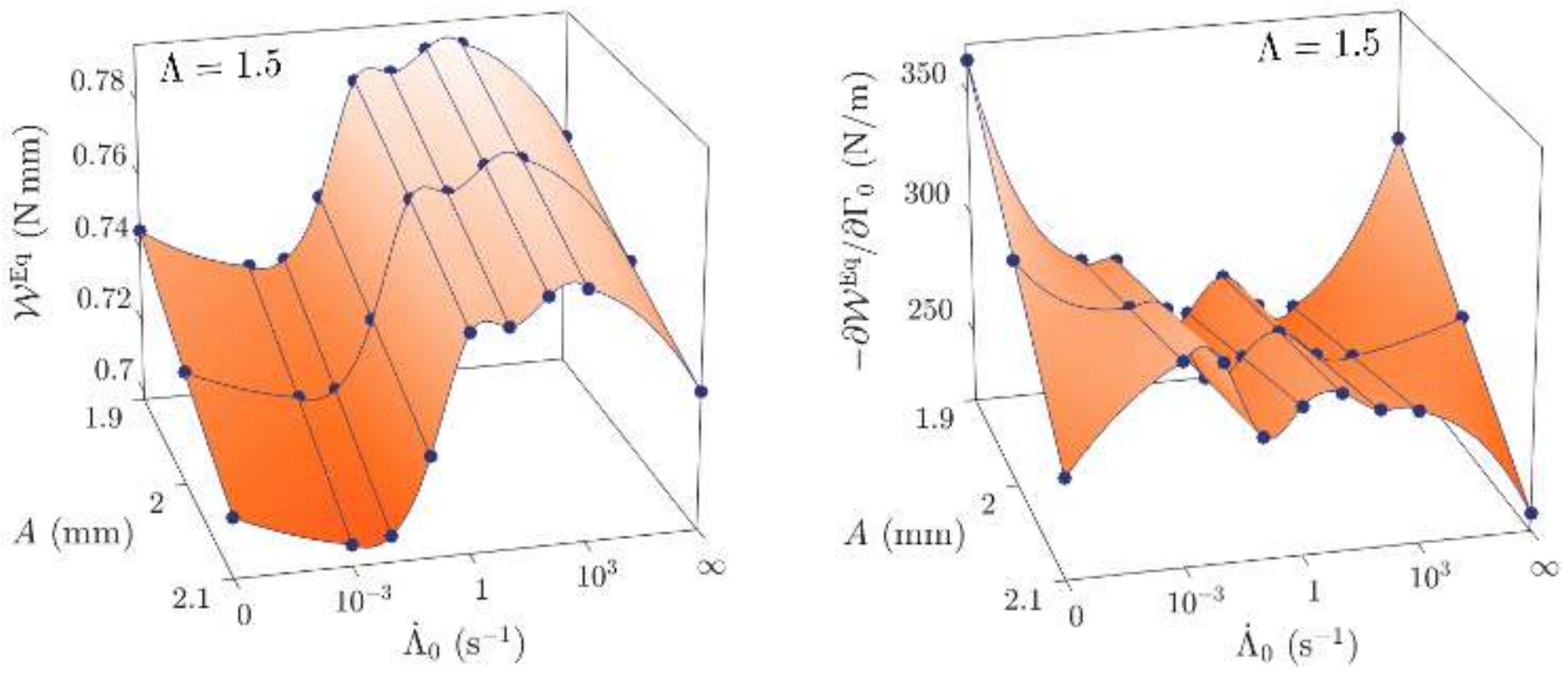

**Figure 4** Representative examples of the equilibrium energy (37) and its derivative (38) plotted as functions of the global stretch rate $\dot{\Lambda}_0$ and the initial crack size $A$ at a fixed global stretch $\Lambda$.

To illustrate the evaluation of the Griffith criticality condition (39), Figure 5 includes a plane of constant value—in this case, $G_c = 150$ N/m—whose intersection with the surface of the derivative $-\partial\mathcal{W}^{\mathrm{Eq}}/\partial\Gamma_0$ allows to readily identify the critical global stretches $\Lambda_c$ at which the crack grows for a given critical energy release rate $G_c$ as a function of the global stretch rate $\dot{\Lambda}_0$.

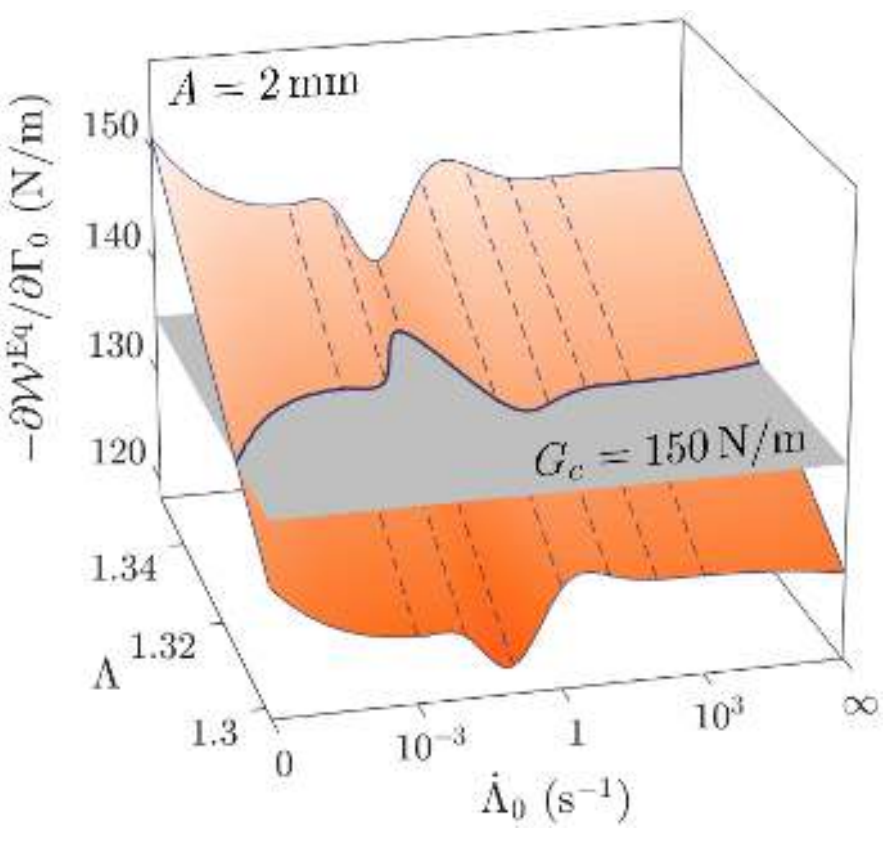

**Figure 5** Representative example of the derivative (38) of the equilibrium elastic energy plotted as a function of the global stretch $\Lambda$ and the global stretch rate $\dot{\Lambda}_0$ for a fixed initial crack size $A$.

# 4 A comprehensive parametric study of the effects of the elasticity and viscosity of the elastomer and of the 3D geometry of the cracks and specimens

At this stage, we are in a position to deploy the Griffith criticality condition (2) to analyze the single edge notch fracture test fully accounting for the viscoelasticity of elastomers and the 3D geometry of the pre-existing cracks and the specimens.

For definiteness, in the parametric study that follows, we consider specimens of length $L = 15\,\mathrm{mm}$, width $H = 4\,\mathrm{mm}$, and thickness $B = 0.5\,\mathrm{mm}$ that contain a pre-existing crack of three different lengths $A = 1.9$, 2, and 2.1 mm. These specific values for $L$, $H$, $B$, and $A$ are chosen here because they are representative of those used in experiments (Hamed and Park 1999; Cristiano et al. 2011; Fan et al. 2017; Roucou et al. 2018; Slootman et al. 2020; Le Menn 2022; Wang and Fan 2023; Xue et al. 2024).

As also representative of the behavior of typical elastomers (Lopez-Pamies 2010; Breedlove et al. 2024), we make use of the material constants

$$\mu_1 = 0.09\,\mathrm{MPa}, \quad \alpha_1 = 0.5, \quad \mu_2 = 0.01\,\mathrm{MPa}, \quad \alpha_2 = 2.5 \tag{46}$$

in the equilibrium stored-energy function (24a) that describe the non-Gaussian elasticity of the elastomer at states of thermodynamic equilibrium. Note that $\mu^{\mathrm{Eq}} = \mu_1 + \mu_2 = 0.1\,\mathrm{MPa}$.

With the objective of thoroughly probing the effects of the additional non-Gaussian elasticity that arises at non-equilibrium states, we consider the three different sets of material constants

$$\begin{cases} \texttt{Weaker Growth:} & \nu_1 = 0.9\nu,\ \beta_1 = 0.5 \quad \text{and} \quad \nu_2 = 0.1\nu,\ \beta_2 = 1, & (47\text{a}) \\ \texttt{Equal Growth:} & \nu_1 = 0.9\nu,\ \beta_1 = 0.5 \quad \text{and} \quad \nu_2 = 0.1\nu,\ \beta_2 = 2.5, & (47\text{b}) \\ \texttt{Stronger Growth:} & \nu_1 = 0.9\nu,\ \beta_1 = 0.5 \quad \text{and} \quad \nu_2 = 0.1\nu,\ \beta_2 = 7.5. & (47\text{c}) \end{cases}$$

with

$$\nu = \nu_1 + \nu_2 = \mu^{\mathrm{NEq}} = 0.5, 1, 2\ \mathrm{MPa} \tag{48}$$

in the non-equilibrium stored-energy function (24b) . As labeled in (47), these material constants yield non-equilibrium stored-energy functions (24b) with weaker ($\beta_2 < \alpha_2$), equal ($\beta_2 = \alpha_2$), and stronger ($\beta_2 > \alpha_2$) growth conditions—in the limit of large deformations as $||\mathbf{F}|| = \sqrt{I_1} \nearrow +\infty$—than that of the equilibrium stored-energy function (24a) with (46).

So as to also thoroughly probe the effects of the value of the viscosity and its nonlinearity, we consider the three sets of material constants

$$\begin{cases} \texttt{Constant Viscosity:}\ N_0 = \eta_0,\ \eta_\infty = \eta_0 & (49\text{a}) \\ \texttt{Shear-thinning Viscosity:}\ N_0 = \eta_0,\ \eta_\infty = 0.01\eta_0,\ K_2 = 500\,\mathrm{MPa}^{-2},\ \gamma_2 = 1 & (49\text{b}) \\ \texttt{Deformation-dependent Viscosity:}\ N_0 = 100\eta_0,\ K_1 = 2,\ \gamma_1 = 4,\ K_2 = 0, & (49\text{c}) \end{cases}$$

with

$$\eta_0 = \tau_0 \nu = 0.5, 1, 2\,\text{MPa s}, \quad \tau_0 = 1\,\text{s}, \tag{50}$$

in the viscosity function (27). As labeled in (49), these values serve to examine the basic case of constant viscosity, as well as those of shear-thinning and deformation-dependent viscosities, one at a time. Note that the value (50) of the initial viscosity $\eta_0$ is chosen in a normalized manner so that the resulting initial relaxation time is kept constant at $\tau_0 = \eta_0/\mu^{\text{NEq}} = 1$ s for all the non-equilibrium initial shear moduli (48).

For the above material constants and pre-existing crack and specimen geometries, the entire range of viscoelastic behaviors—from elasticity-dominated to viscosity-dominated—occurs effectively within the range $\dot{\Lambda}_0 = [10^{-3}, 10^3]$ s$^{-1}$ of global stretch rates. Accordingly, in the parametric study presented in this section, we carry out simulations for the nine global stretch rates

$$\dot{\Lambda}_0 = 0+,\ 10^{-3},\ 10^{-2},\ 10^{-1},\ 10^{0},\ 10^{1},\ 10^{2},\ 10^{3},\ +\infty\,\text{s}^{-1}, \tag{51}$$

which include the purely elastic limit of infinitesimally slow loading, $\dot{\Lambda}_0 = 0+$, as well as the "pseudo-elastic" limit of infinitely fast loading, $\dot{\Lambda}_0 = +\infty$.

In all, there are $3 \times 3 \times 3 = 27$ combinations of material constants, three crack lengths, and nine different stretch rates, which result in a total of 729 simulations. In the following subsections, we present the main findings from these simulations, organized in terms of the types of viscosity of the elastomer and, within this, in terms of the growth conditions of its non-equilibrium elasticity and loading rate.

## 4.1 Constant viscosity

We begin by presenting results that illustrate how the critical values $\Lambda_c$ and $S_c$ of the global stretch $\Lambda$ and the global stress $S$, as well as the corresponding critical tearing energy $T_c$, at which the crack grows in single edge notch fracture tests depend on the loading rate and on the growth conditions of the non-equilibrium elasticity for the basic case when the viscosity of the elastomer is constant.

Figure 6 presents results for the global stress $S = P/(HB)$ as a function of the applied global stretch $\Lambda = l/L$ for specimens stretched at the constant global stretch rates $\dot{\Lambda}_0 = 10^{-3}, 10^{-1}$, $10^0$, $10^1$, $10^3$ s$^{-1}$. The results pertain to specimens with initial crack size $A = 2$ mm, elastomers with non-equilibrium elasticity of weaker, equal, and stronger growth conditions (47), with $\nu = \mu^{\text{NEq}} = 20\mu^{\text{Eq}} = 2$ MPa, than their equilibrium elasticity, and constant viscosity (49a) with $\eta_0 = \nu\tau_0 = 2$ MPa s. All the results are plotted up to the critical global stretch $\Lambda_c$ at which the Griffith criticality condition (39) is satisfied (indicated by an empty circle in all the force-deformation plots) for the representative case when the critical energy release rate is $G_c = 150$ N/m.

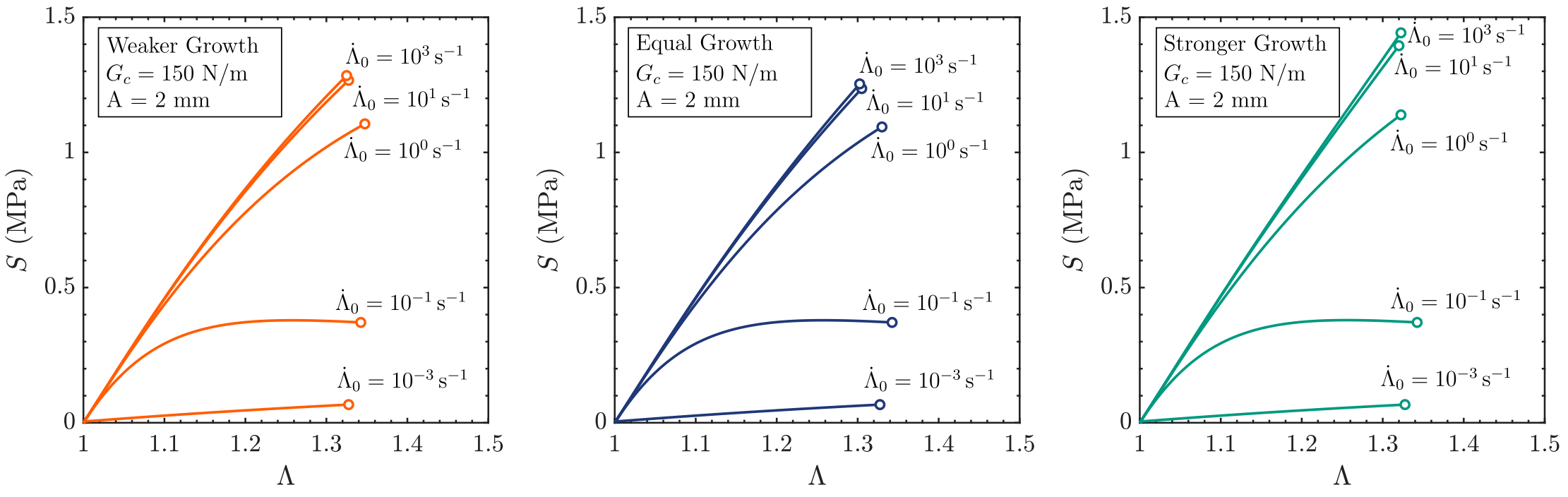

**Figure 6** Force-deformation response of specimens with pre-existing cracks of length $A = 2$ mm stretched at various constant global stretch rates $\dot{\Lambda}_0$. The results pertain to elastomers with constant viscosity (49a) and non-equilibrium elasticity of weaker (47a), equal (47b), and stronger (47c) growth conditions than their equilibrium elasticity.

An immediate observation from the results in Figure 6 is that the critical global stretches $\Lambda_c$ and stresses $S_c$ at which the crack grows are dependent on the loading rate $\dot{\Lambda}_0$ and on the growth conditions of the non-equilibrium elasticity of the elastomer, as characterized by the material constant $\beta_2$. Interestingly, the dependence of $\Lambda_c$ on $\dot{\Lambda}_0$ and $\beta_2$ is non-monotonic, while that of $S_c$ is monotonic on $\dot{\Lambda}_0$ but not so on $\beta_2$.

The quantitative dependence of $\Lambda_c$ and $S_c$ on $\dot{\Lambda}_0$ and $\beta_2$ are better illustrated by Figure 7. The results for the stretch show that, save for the special case of equal growth ($\beta_2 = \alpha_2 = 2.5$), the smallest $\Lambda_c$ is attained at the slowest loading rate $\dot{\Lambda}_0 = 0+$, when the behavior of the material is purely elastic. As the loading rate $\dot{\Lambda}_0$ increases so does $\Lambda_c$, but only until certain threshold $\dot{\Lambda}_0^{\text{th}}$ at which $\Lambda_c$ reaches a local maximum. Further increase in $\dot{\Lambda}_0$ beyond that threshold leads to the evolution of $\Lambda_c$ towards an asymptotic value. The threshold $\dot{\Lambda}_0^{\text{th}}$ at which $\Lambda_c$ reaches its local maximum appears to correlate with the initial relaxation time $\tau_0 = \eta_0/\mu^{\text{NEq}}$, roughly, $\dot{\Lambda}_0^{\text{th}} \approx \tau_0^{-1}$. On the other hand, the results for the stress show that $S_c$ increases monotonically with increasing loading rate $\dot{\Lambda}_0$, this irrespective of the value of $\beta_2$. The dependence of both $\Lambda_c$ and $S_c$ on the non-equilibrium elasticity of the elastomer is clearly non-monotonic and intricately coupled with the loading rate.

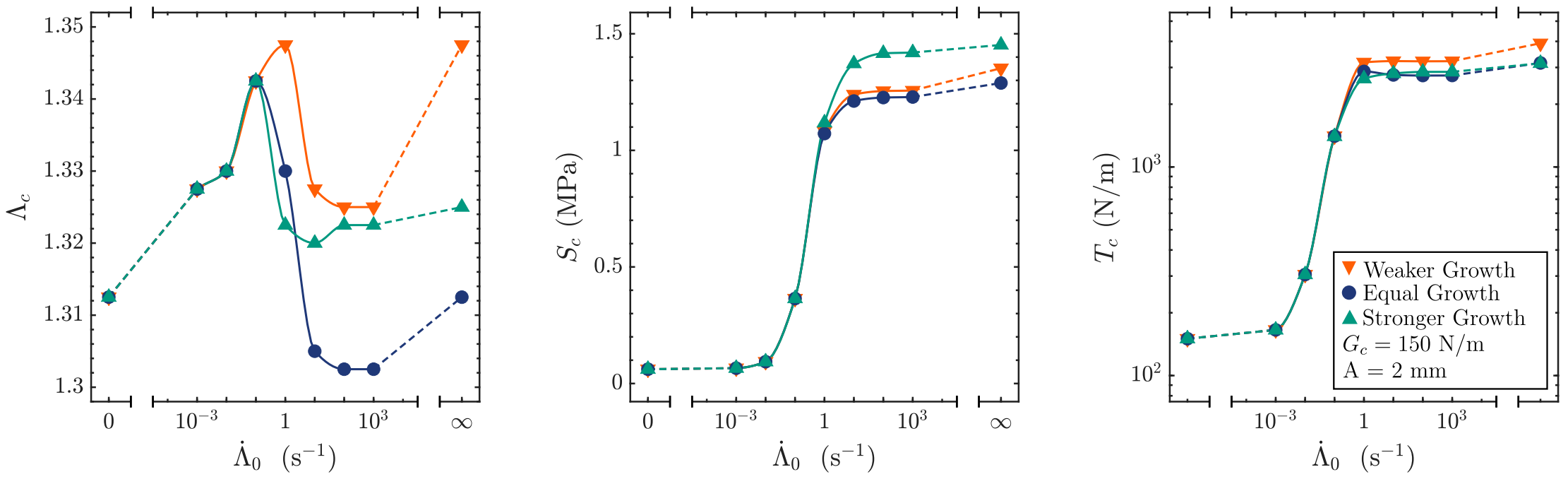

**Figure 7** The critical global stretch $\Lambda_c$, global stress $S_c$, and tearing energy $T_c$ at which the crack grows plotted as functions of the global stretch rate $\dot{\Lambda}_0$ at which the test is carried out. The results pertain to elastomers with constant viscosity (49a).

In addition to the results for the critical global stretch $\Lambda_c$ and the critical global stress $S_c$, it is instructive to plot the results for the corresponding critical tearing energy (44). Figure 7 includes such a plot in terms of the global stretch rate $\dot{\Lambda}_0$ for the three material constants $\beta_2 = 1, 2.5, 7.5$. The results show the typical "S" shape found in other types of fracture tests, such as the trousers fracture test (Greensmith and Thomas 1955), with the caveat that they are not strictly monotonic. Specifically, as $\dot{\Lambda}_0 \searrow 0$, for sufficiently slow global stretch rates, $T_c \searrow G_c$. As $\dot{\Lambda}_0$ increases, so does $T_c$, although not necessarily monotonically. As $\dot{\Lambda}_0 \nearrow +\infty$, for sufficiently fast global stretch rates, $T_c$ approaches an asymptotic value, $T_c^\infty$ say. The transition of $T_c$ from its minimum value $G_c$ to its asymptotic value $T_c^\infty$ is controlled by both the non-equilibrium elasticity of the elastomer and its viscosity. Given that $G_c = 150\,\text{N/m}$ and $T_c^\infty > 3000\,\text{N/m}$, the results in Figure 7 make it plain that the critical tearing energy $T_c$ is primarily a direct manifestation of the viscoelastic behavior of the elastomer, and *not* of its fracture behavior. In other words, in view of the representation (5), the value of $T_c$ is mostly a measure of how much energy is dissipated by viscous deformation and *not* by the creation of surface.

## 4.2 Shear-thinning viscosity

As already pointed out above, elastomers are by and large characterized *not* by a constant viscosity but by a strongly nonlinear viscosity that depends both on the deformation and on the deformation rate. In this subsection, we examine the effects of the latter on $\Lambda_c$, $S_c$, and $T_c$.

Figure 8 presents results analogous to those presented in Figure 7 for elastomers whose viscosity is *not* constant but instead decreases with deformation rate, that is, a viscosity that exhibits shear thinning; the corresponding force-deformation results analogous to those in Figure 6 are presented in Figure A.2. Precisely, the results pertain to specimens with initial crack size $A = 2\,\text{mm}$, elastomers with non-equilibrium elasticity of weaker, equal, and stronger growth conditions (47), with $\nu = \mu^{\rm NEq} = 20\mu^{\rm Eq} = 2$ MPa, than their equilibrium elasticity, and shear-thinning viscosity (49b) with $\eta_0 = \nu\tau_0 = 2$ MPa s. Much like the results in Figure 7, the results in Figure 8 correspond to the representative case when the critical energy release rate is $G_c = 150$ N/m.

A quick glance suffices to recognize that the results in Figure 8 for shear-thinning viscosity are qualitatively similar to those in Figure 7 for constant viscosity. Quantitatively, Figure 8 shows that the presence of shear thinning shifts the threshold $\dot{\Lambda}_0^{\rm th}$ at which $\Lambda_c$ reaches a local maximum to larger values. Shear thinning leads as well to larger differences in the values that $\Lambda_c$ can take, significantly more so than the differences that result from variations in the growth conditions of the non-equilibrium elasticity of elastomers with constant viscosity. For instance, note that, roughly, $\Lambda_c = 1.35$ at $\dot{\Lambda}_0 = 10^{-3}$ s$^{-1}$ while $\Lambda_c = 2.6$ at $\dot{\Lambda}_0 = 10^1$ s$^{-1}$ for all growth conditions. By contrast, the presence of shear thinning appears to have a more moderate quantitative effect on the critical global stress $S_c$ and critical tearing energy $T_c$.

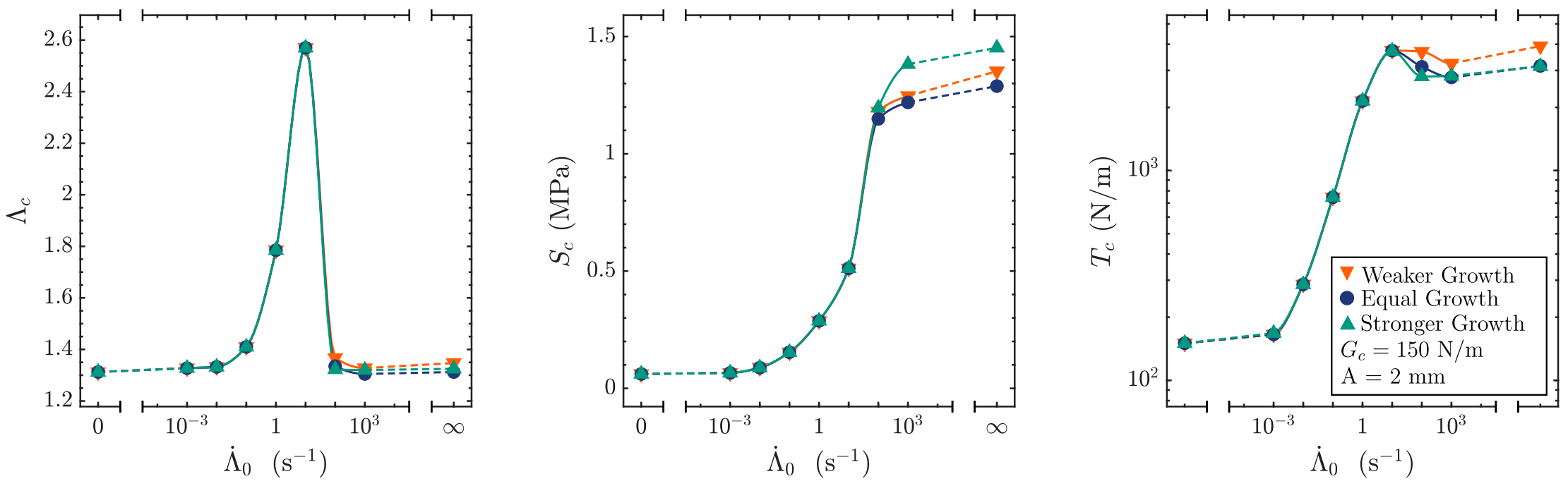

**Figure 8** The critical global stretch $\Lambda_c$, global stress $S_c$, and tearing energy $T_c$ at which the crack grows plotted as functions of the global stretch rate $\dot{\Lambda}_0$ at which the test is carried out. The results pertain to elastomers with shear-thinning viscosity (49b).

## 4.3 Deformation-dependent viscosity

Next, we examine the effects that a nonlinear viscosity that depends on deformation can have on $\Lambda_c$, $S_c$, and $T_c$.

Figure 9 presents results analogous to those presented in Figures 7 and 8 for elastomers whose viscosity increases with increasing deformation; the corresponding force-deformation results analogous to those in Figure 6 are presented in Figure A.3. In particular, the results pertain to specimens with initial crack size $A = 2$ mm, elastomers with non-equilibrium elasticity of weaker, equal, and stronger growth conditions (47), with $\nu = \mu^{\rm NEq} = 20\mu^{\rm Eq} = 2$ MPa, than their equilibrium elasticity, and deformation-dependent viscosity (49c) with $\eta_0 = \nu\tau_0 = 2$ MPa s. Again, much like the results in the two preceding subsections, the results in Figure 9 correspond to the representative case when the critical energy release rate is $G_c = 150$ N/m.

Qualitatively, the results in Figure 9 for a nonlinear viscosity that increases with deformation are similar to those in Figure 7 for constant viscosity. Quantitatively, two observations are called for when comparing Figure 9 with Figure 7. First, the presence of a nonlinear viscosity that increases with deformation shifts the threshold $\dot{\Lambda}_0^{\rm th}$ at which $\Lambda_c$ reaches a local maximum to smaller values. Second, on its own, contrary to a shear-thinning viscosity, a nonlinear viscosity that increases with deformation does not result in a range of values for $\Lambda_c$ that are substantially different from those that result from a constant viscosity. The same is true for $S_c$ and $T_c$.

## 4.4 The classical formula (6) of Rivlin and Thomas vs. 3D full-field results

The classical formula (6) of Rivlin and Thomas (1953)—together with the approximation (7) for the numerical factor $K(\Lambda)$ that emerged from the works of Greensmith (1963) and Lindley (1972)—is commonly used well beyond its original scope, see Section 1.1, in the experimental literature so as to estimate the values of the critical tearing energy $T_c$ from single edge notch fracture tests. In the sequel, we show how such an overextended use can lead to significant errors in the estimation of $T_c$. We do so by confronting the formula (6) directly with an exact solution from our 3D full-field simulations for viscoelastic elastomers.

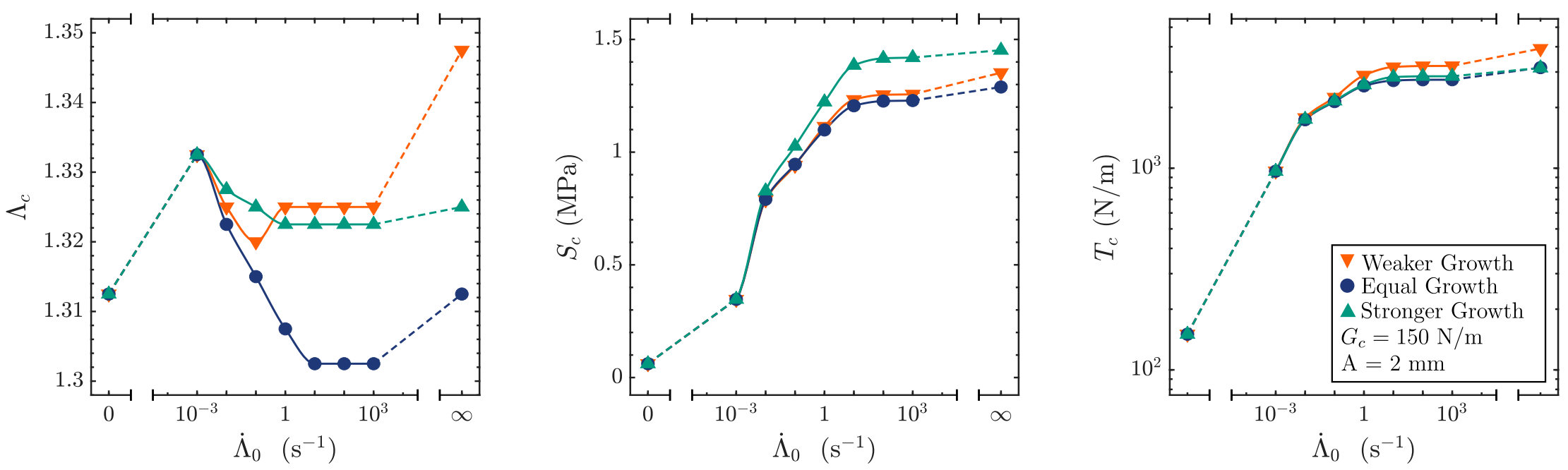

**Figure 9** The critical global stretch $\Lambda_c$, global stress $S_c$, and tearing energy $T_c$ at which the crack grows plotted as functions of the global stretch rate $\dot{\Lambda}_0$ at which the test is carried out. The results pertain to elastomers with deformation-dependent viscosity (49c).

**Remark 4** In order to use the formula (6) for elastomers that are not purely elastic solids but solids that dissipate energy by deformation, such as viscoelastic elastomers, the stored-energy function $W_{\rm un}(\Lambda)$ is typically reinterpreted as the "strain energy density" (stored and dissipated) of the elastomer when subjected to a uniaxial tension test that is carried out at the same stretch rate $\dot{\Lambda}_0$ as the single edge notch fracture test of interest.

Figure 10 provides a comparison between the energy release rate $-\partial\mathcal{W}/\partial\Gamma_0$ generated by the classical formula (6), with the numerical factor (7), and the exact result (44) determined from 3D full-field simulations of a single edge notch test, carried out at the global stretch rate $\dot{\Lambda}_0 = 10^0$ s$^{-1}$, for a specimen with pre-existing crack of length $A = 2$ mm and an elastomer with shear-thinning viscosity (49b), $\eta_0 = \nu\tau_0 = 2$ MPa s, and non-equilibrium elasticity of equal growth conditions (47b), with $\nu = \mu^{\rm NEq} = 20\mu^{\rm Eq} = 2$ MPa, than its equilibrium elasticity.

**Figure 10** Representative comparison between the energy release rate generated by the classical Rivlin-Thomas formula (6), with the numerical factor (7), and the exact result (44) determined from 3D full-field simulations of a single edge notch test carried out at the global stretch rate $\dot{\Lambda}_0 = 10^0$ s$^{-1}$. The results pertain to a specimen with pre-existing crack of length $A = 2$ mm and an elastomer with shear-thinning viscosity (49b) and non-equilibrium elasticity of equal growth conditions (47b) to its equilibrium elasticity.

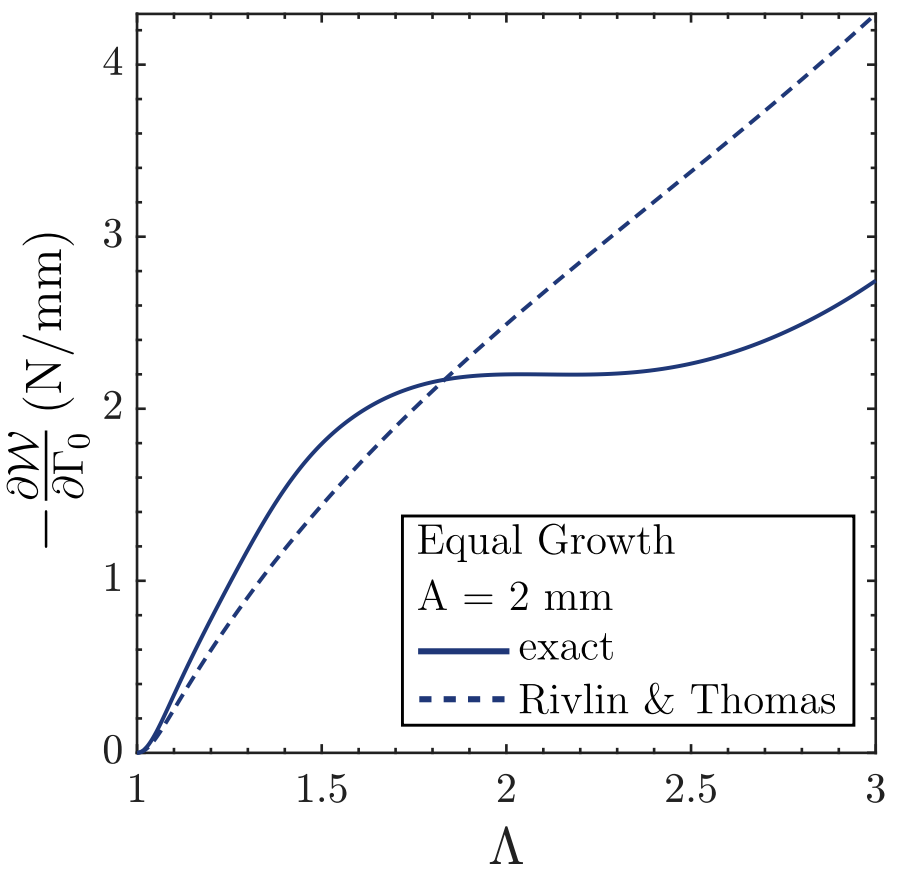

The comparison in Figure 10 makes it clear that the classical formula (6) of Rivlin and Thomas (1953) can be greatly inaccurate, increasingly more so for larger global stretches $\Lambda$, when used for elastomers that dissipate energy by deformation and for cracks that are not much smaller than the length and the width of the specimen.

## 5 Comparisons with experiments on an acrylate elastomer

As alluded to in the Introduction, since the pioneering work of Busse (1934) in the 1930s, rather remarkably, only a handful of the numerous single edge notch fracture tests that have been reported in the literature have been carried out at different loading rates $\dot{\Lambda}_0$. In this section, we deploy the Griffith criticality condition (2) to explain one of them, those of Le Menn (2022) on an acrylate elastomer.

As part of their doctoral thesis, Le Menn (2022) carried out a series of single edge notch tests on dog-bone specimens of gauge length $L = 15$ mm, width $H = 4$ mm, thickness $B \in [0.2, 0.8]$ mm, and crack size $A = 1$ mm made of acrylate elastomers (for use as coatings of optical fibers) at room temperature and four different constant global stretch rates, approximately given by $\dot{\Lambda}_0 = 6.7 \times 10^{-5}, 6.7 \times 10^{-4}, 6.7 \times 10^{-3}, 3.3 \times 10^{-2}$ s$^{-1}$; see Appendix B.4 in their work. Out of all the various compositions that Le Menn (2022) considered, we focus here on the one that she labeled OL HEA 30% PEA for which she provided experimental data for the elastic response of the elastomer in the form of stress-stretch data for uniaxial and equi-biaxial tension carried out at a constant stretch rate of about $\dot{\lambda}_0 = 6.7 \times 10^{-3}$ s$^{-1}$, sufficiently slow that viscous effects were not observed; see Chapter 3 in their work.

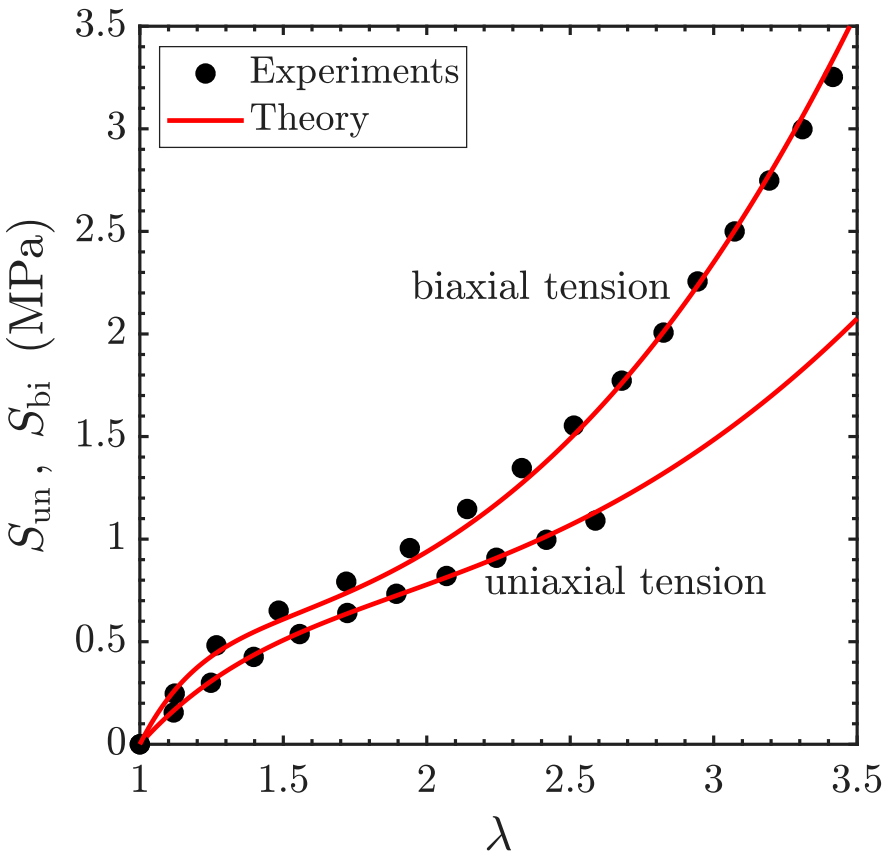

**Figure 11** Comparison between the stress-stretch response described by the viscoelastic model (25)-(26), with the material constants in Table 1, and the experimental data reported by Le Menn (2022) for the acrylate elastomer OL HEA 30% PEA subjected to uniaxial and equi-biaxial tension applied at a slow constant stretch rate $\dot{\lambda}_0$ at room temperature.

**The viscoelastic behavior.** The fitting of the material constants $\mu_1, \mu_2, \alpha_1, \alpha_2$, which, again, describe the equilibrium elasticity in the viscoelastic model (25)-(26), to the uniaxial and equi-biaxial data reported by Le Menn (2022) yields the values listed in the first column of Table 1. As seen from the comparisons presented in Figure 11, the constitutive relation (25)-(26) with such material constants describes reasonably well the experimentally measured response at states of thermodynamic equilibrium.

**Table 1** Values of the material constants in the viscoelastic model (25)-(26) with (52) for the acrylate elastomer OL HEA 30% PEA.

| | | |
|---|---|---|
| $\mu_1 = 0.400$ MPa | $\nu_1 = 5.19$ MPa | $\eta_0 = 80$ MPa s |
| $\alpha_1 = 0.0488$ | $\beta_1 = -0.5$ | $\eta_\infty = 0$ |
| $\mu_2 = 0.119$ MPa | $\nu_2 = 0$ | $K_2 = 0.01$ MPa$^{-2}$ |
| $\alpha_2 = 1.9985$ | $\beta_2 = 0$ | $\gamma_2 = 1$ |

It remains to determine the materials constants $\nu_1, \nu_2, \beta_1, \beta_2$ for the non-equilibrium elasticity and the viscosity function $\eta(\mathbf{C}, \mathbf{C}^v)$. Given the lack of direct experimental data beyond that presented in Figure 11 for the viscoelastic behavior of OL HEA 30% PEA, we turn to remarking that the force-deformation response of the specimens in the single edge notch fracture tests reported by Le Menn (2022), reproduced in Figure 12(a), differ only on the value of the critical global stretch $\Lambda_c$—and hence also on the value of the critical global stress $S_c$—at which fracture nucleates. In other words, the force-deformation response of the specimens appears to be independent of the loading rate $\dot{\Lambda}_0$, except for the fact that $\Lambda_c$ and $S_c$ increase with increasing $\dot{\Lambda}_0$. Given that the stretch and stretch rates around the crack front in single edge notch fracture tests are locally very large, such a behavior may be attributed to several plausible mechanisms. One possibility is that it is a direct manifestation of the strength surface of the elastomer; see, e.g. (Kamarei et al. 2025, Sections 2.2.1 and 5.1). Another possibility is that the viscosity of the elastomer is one that is strongly nonlinear at large stretches and stretch rates, but negligible otherwise. In this work, we explore this latter mechanism. Accordingly, we consider a viscosity function of the piecewise form

$$\eta(\mathbf{C}, \mathbf{C}^v) = \begin{cases} \eta_0, & \mathbf{X} \in \Omega_0^{\rm I} \\ \eta_\infty + \dfrac{\eta_0 - \eta_\infty}{1 + (K_2 \mathcal{J}_2^{\rm NEq})^{\gamma_2}}, & \mathbf{X} \in \Omega_0^{\rm II} \\ \eta_\infty, & \mathbf{X} \in \Omega_0 \setminus (\Omega_0^{\rm I} \cup \Omega_0^{\rm II}) \end{cases} \quad (52)$$

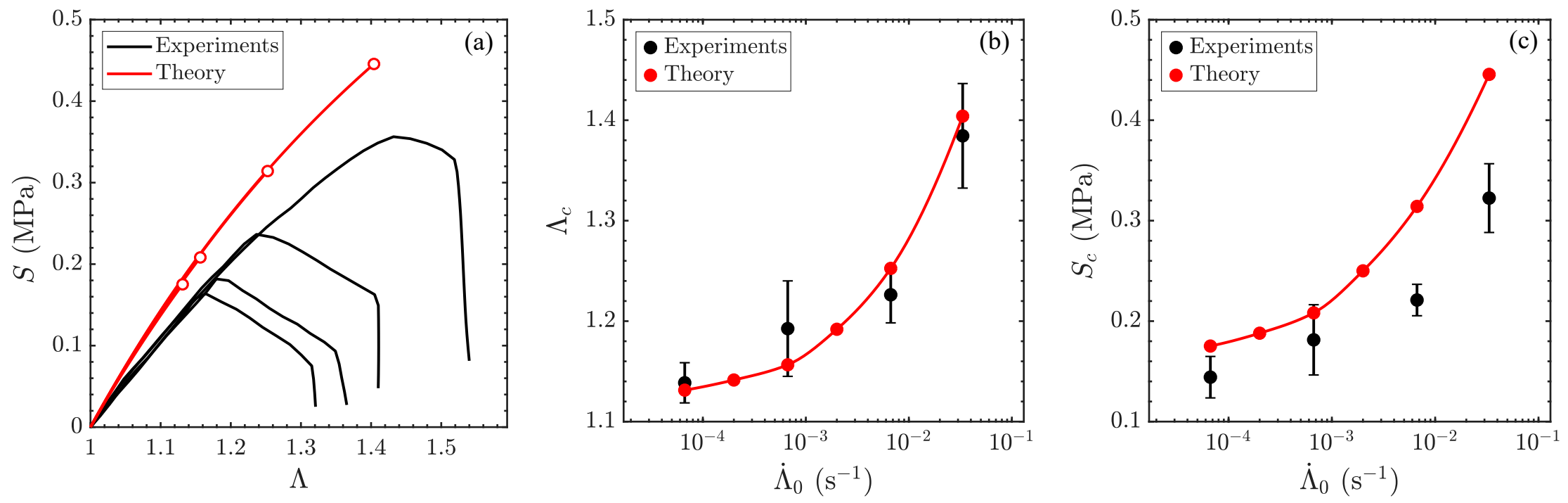

**Figure 12** Comparison with experiments. (a) Comparison between the force-deformation response predicted theoretically and the experimental data reported by Le Menn (2022) for single edge notch fracture tests on the acrylate elastomer OL HEA 30% PEA carried out at four different global stretch rates $\dot{\Lambda}_0$. (b), (c) Corresponding comparison between the critical global stretch $\Lambda_c$ and the critical global stress $S_c$ at which fracture nucleates predicted by the Griffith criticality condition (2) and the experimental results.

where, for definiteness, $\Omega_0^{\rm I} = \{\mathbf{X} : |X_1| \leqslant B/2, (X_2-X_2^{\rm I})^2+X_3^2 < (0.05A)^2\}$ and $\Omega_0^{\rm II} = (\{\mathbf{X} : |X_1| \leqslant B/2, (X_2-X_2^{\rm II})^2+X_3^2 < (0.7A)^2\}) \setminus \Omega_0^{\rm I}$ with $X_2^{\rm I} = 1.05\,{\rm A} - 0.05\,{\rm mm}$ and $X_2^{\rm II} = 1.7\,{\rm A} - 0.05\,{\rm mm}$. As illustrated by Figure 13, $\Omega_0^{\rm I}$ stands for a small region surrounding the crack front where the viscosity is assumed to be of constant value $\eta_0$. In such a region, the stretch and stretch rates are extremely large and we assume that whatever shear-thinning of the viscosity may exist, it is canceled out by its enhancement due to deformation. Furthermore, $\Omega_0^{\rm II}$ stands for a larger but still relatively small enclosing region where the viscosity is assumed to decrease from $\eta_0$ to $\eta_\infty \ll \eta_0$ due to shear thinning. Outside $\Omega_0^{\rm I}$ and $\Omega_0^{\rm II}$, in the remaining of the specimen, the viscosity is assumed to be of constant value $\eta_\infty$.

Given the viscosity function (52), a search of the material constants $\nu_1$, $\nu_2$, $\beta_1$, $\beta_2$, $\eta_0$, $\eta_\infty$, $K_2$, $\gamma_2$ in the viscoelastic model (25)-(26) that are consistent with the data of Le Menn (2022) yields the values listed in the second and third columns of Table 1.

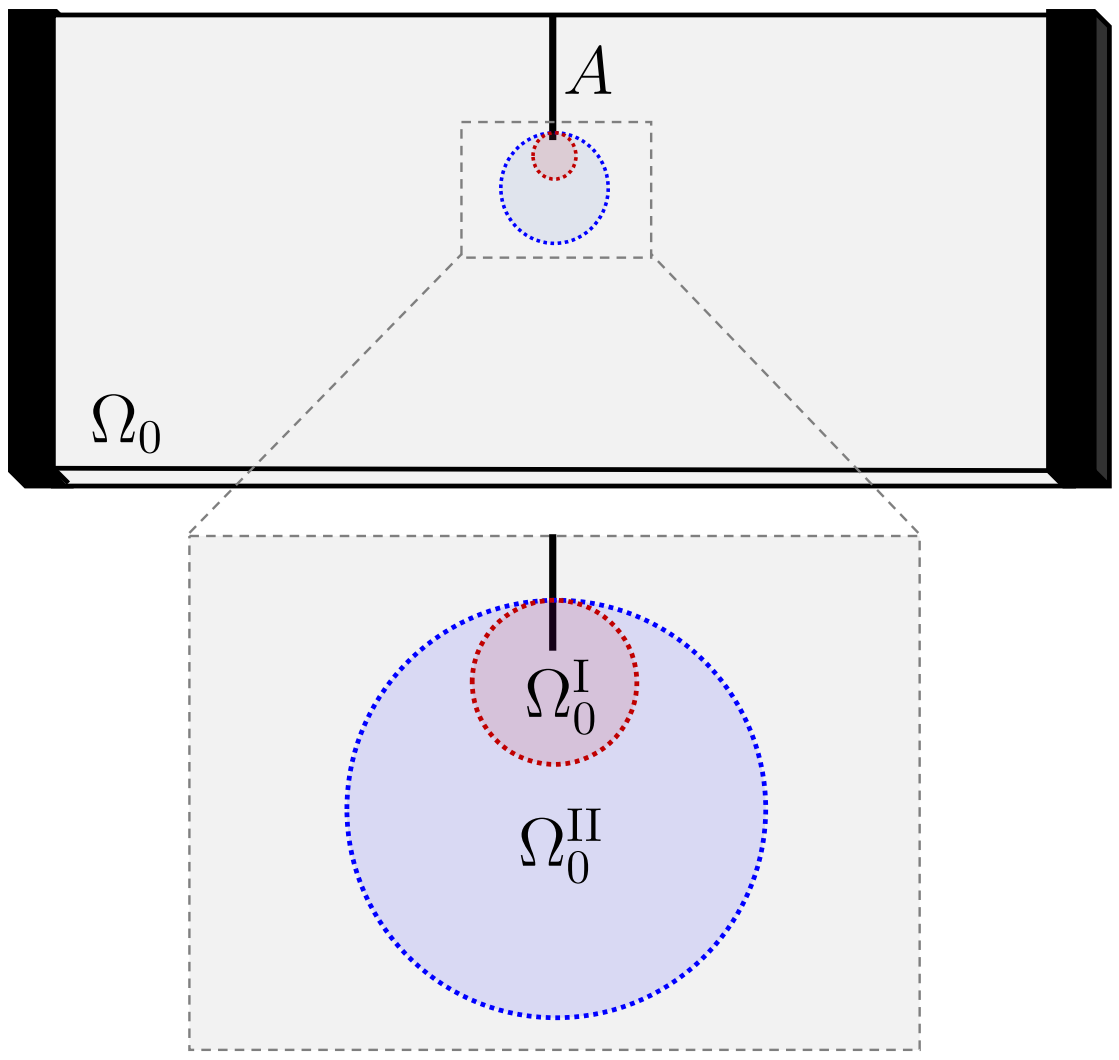

**Figure 13** Schematics of the small regions $\Omega_0^{\rm I}$ and $\Omega_0^{\rm II}$ around the crack front in the single edge notch fracture tests carried out by Le Menn (2022) wherein the viscoelastic behavior of the acrylate elastomer OL HEA 30% PEA is described by the nonlinear viscosity $(52)_{1,2}$.

**The critical energy release rate.** In addition to the viscoelastic behavior of the elastomer, the use of the Griffith criticality condition (2) requires knowledge of its critical energy release rate $G_c$. For the problem at hand, the fitting of $G_c$ in (2) from the single edge notch fracture test carried out at the slowest loading rate $\dot{\Lambda}_0 = 6.7\times10^{-5}$ s$^{-1}$ yields

$$G_c = 92\,{\rm N/m}, \tag{53}$$

which is consistent with the estimate provided in (Le Menn 2022, Table 3.7).

At this stage, having calibrated the viscoelastic behavior and the critical energy release rate of the elastomer, we are ready to deploy the Griffith criticality condition (2) to confront its predictions with the single edge notch fracture tests reported by Le Menn (2022). Figures 12(b) and 12(c) do so for the critical global stretch $\Lambda_c$ and the critical global stress $S_c$ at which fracture nucleates as functions of the applied global stretch rate $\dot{\Lambda}_0$. The error bars for the experimental data correspond to the neighborhood over which the force-deformation response of the specimens approaches its peak value and hence over which the onset of fracture is bound to occur.

The main observation from Figures 12(b) and 12(c) is that the Griffith criticality condition (2) is in good overall agreement with the experimental observations. There is an appreciable quantitative difference between the theoretically predicted values for the critical global stress $S_c$ and those measured experimentally. This difference is consistent with that shown by Figure 12(a) between the force-deformation response of the specimens predicted by the theory and that measured experimentally. To wit, the response in the experiments is softer from the outset. This softer behavior may be due to the cracks in the experiments being actually larger than $A = 1$ mm and/or to some experimental error in the measurement of the global stretch $\Lambda$. Howbeit, both sets of results exhibit essentially the same behavior in the force-deformation response and in the critical global stretch $\Lambda_c$ and critical global stress $S_c$ at fracture.

## 6 Summary and final comments

The parametric study presented in Section 4 and the comparisons with experiments presented in Section 5 have shown that the nonlinear viscoelastic properties of elastomers have a profound effect on their behavior in single edge notch fracture tests. Precisely, the critical global stretch $\Lambda_c$ and the critical global stress $S_c$ at which the pre-existing crack in such tests starts to grow depend on the non-Gaussian elasticity of the elastomer at hand and, even more so, on the nonlinearity of its viscosity. For tests carried out at constant global stretch rates $\dot{\Lambda}_0$, the smallest values of the critical global stretch $\Lambda_c$ and the critical global stress $S_c$ are typically attained at infinitesimally slow loading rates $\dot{\Lambda}_0 = 0+$, when viscous effects are absent. As the loading rate increases, so does $\Lambda_c$, but not necessarily in a monotonic fashion. On the other hand, $S_c$ appears to always increase monotonically with increasing loading rate. Elastomers with shear-thinning viscosity tend to exhibit the strongest sensitivity to the loading rate.

The strong sensitivity of the results of single edge notch fracture tests to the nonlinear viscoelastic properties of elastomers is due to the fact that the region around the crack front in such tests experiences a wide range of very large stretches and stretch rates—more so than in other standard fracture tests, such as the "pure-shear" and the trousers tests—that fully probes the non-Gaussian elasticity and the nonlinear viscosity of the elastomer being tested. As a corollary of this sensitivity, one can envision using single edge notch fracture tests *not* to measure fracture properties but to indirectly measure viscoelastic properties at large stretches and stretch rates that would be otherwise inaccessible by direct experimental techniques.

Additionally, the parametric study presented in Section 4 has served to establish that the classical formula (6) of Rivlin and Thomas (1953), together with the approximation (7) for the numerical factor $K(\Lambda)$ due to Greensmith (1963) and Lindley (1972), can be greatly inaccurate when used to estimate the critical tearing energy $T_c$ for elastomers that dissipate energy by deformation and for cracks that are not much smaller than the length and the width of the specimen. This result should not come as a surprise, since Rivlin and Thomas (1953) derived the formula (6) precisely under the premise that energy dissipation (viscous or otherwise) by deformation is absent and that the pre-existing crack is much smaller than the length and the width of the specimen, but much larger than its thickness.

The comparisons with experiments presented in Section 5 have also served to add to previous validation results for other tests and materials (Shrimali and Lopez-Pamies 2023a; Lopez-Pamies and Shrimali 2023; Shrimali and Lopez-Pamies 2023b) and thus they have provided further evidence that the Griffith criticality condition (2) may indeed be the condition that governs when large cracks grow in viscoelastic elastomers subjected to arbitrary quasi-static loading conditions.

To conclusively establish whether the Griffith criticality condition (2) is "the law of the land", more comparisons with experiments on different elastomers under different types of loadings would be essential. In this regard, we emphasize that a key implication of (2) is that any rate effects on the nucleation of fracture from large pre-existing cracks in elastomers are due to their nonlinear viscoelastic behavior. This, we hope, will motivate experimentalists and theoreticians alike to thoroughly characterize the nonlinear viscoelasticity of elastomers over large ranges of stretches and stretch rates—and *not* just within the classical limited setting of small deformations, as most often done (Persson et al. 2024), which is insufficient to analyze fracture—so as to help establishing whether all rate effects observed in the fracture of elastomers are indeed of a viscous nature.

# A Additional results on constant, shear-thinning, and deformation-dependent viscosity

Figure A.1 presents results—that complement those presented in Figures 3 to 5 in the main body of the text—for the non-equilibrium stored-energy function $\psi^{\rm NEq}(I_1^e)$ and the dissipated viscous energy density

$$\overline{\phi}(\mathbf{X},t) := \int_0^T 2\phi(\mathbf{F}(\mathbf{X},\tau), \mathbf{C}^v(\mathbf{X},\tau), \dot{\mathbf{C}}^v(\mathbf{X},\tau))\,\mathrm{d}\tau \tag{A.1}$$

over the initial $\Omega_0$ and the current $\Omega(t)$ configurations of a specimen, with pre-existing crack of length $A = 2$ mm, at a global stretch $\Lambda = 1.5$, that has been stretched at the global stretch rate $\dot{\Lambda}_0 = 10^0$ s$^{-1}$. The results pertain to the same single edge notch fracture test considered in Section 3 in the main body of the text, that of an elastomer with material constants $\mu_1 = 0.09$ MPa, $\alpha_1 = 0.5$, $\mu_2 = 0.01$ MPa, $\alpha_2 = 2.5$, $\nu_1 = 1.8$ MPa, $\beta_1 = 0.5$, $\nu_2 = 0.2$ MPa, $\beta_2 = 7.5$, $\eta_0 = N_0 = \eta_\infty = 2$ MPa s, and a specimen of length $L = 15$ mm, width $H = 4$ mm, and thickness $B = 0.5$ mm.

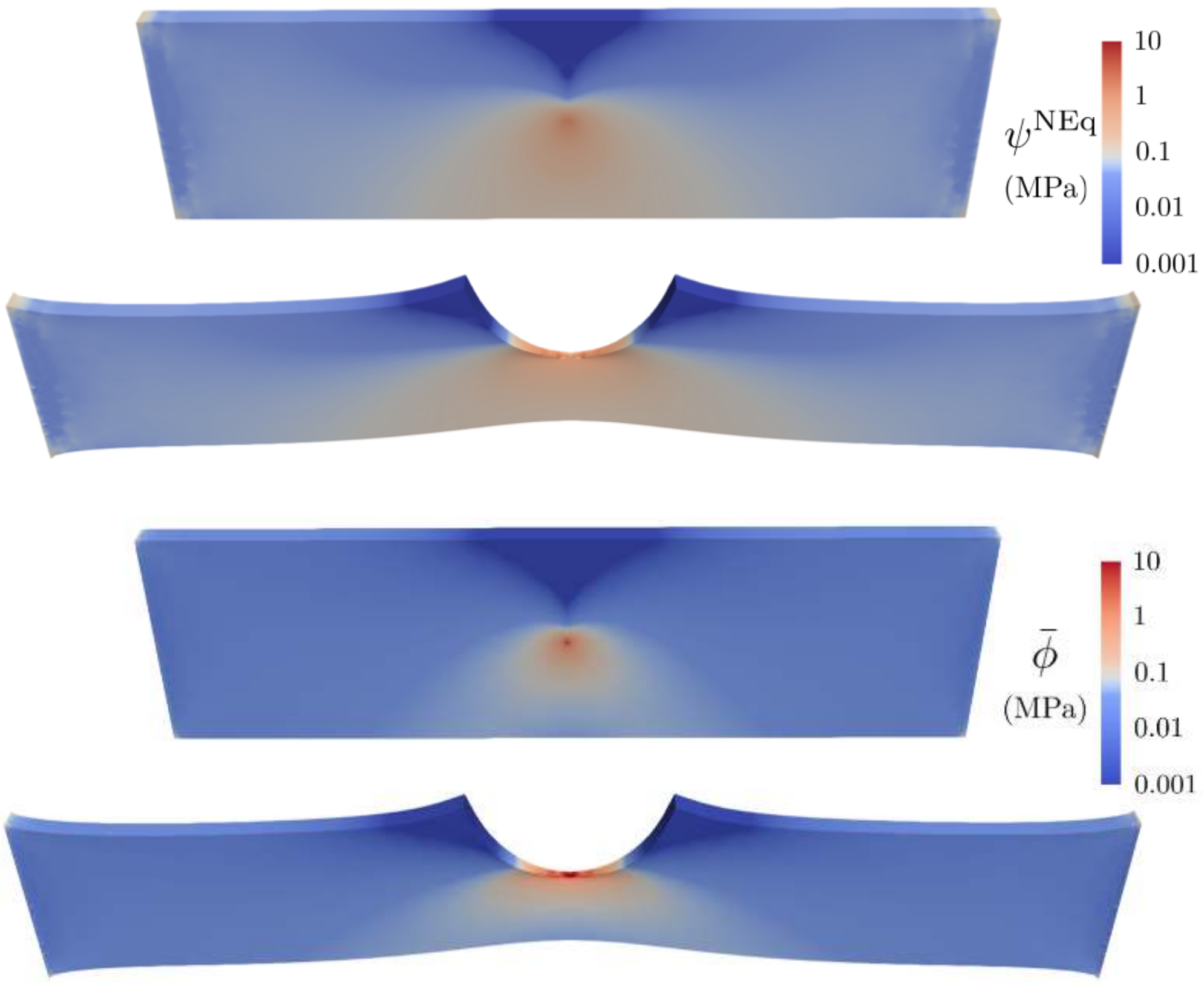

**Figure A.1** Representative FE solution for the non-equilibrium stored-energy function $\psi^{\rm NEq}(I_1^e)$ and the dissipated viscous energy density $\overline{\phi}$ over the initial $\Omega_0$ and current $\Omega(t)$ configurations of a specimen, with pre-existing crack of length $A = 2$ mm, at a global stretch $\Lambda = 1.5$, that has been stretched at the global stretch rate $\dot{\Lambda}_0 = 1\,\mathrm{s}^{-1}$.

Figure A.2 presents results—that complement those presented in Figure 8 in the main body of the text—for the global stress $S = P/(HB)$ as a function of the applied global stretch $\Lambda = l/L$ for specimens stretched at the constant global stretch rates $\dot{\Lambda}_0 = 10^{-3}, 10^{-1}, 10^{0}, 10^{1}, 10^{3}$ s$^{-1}$. The

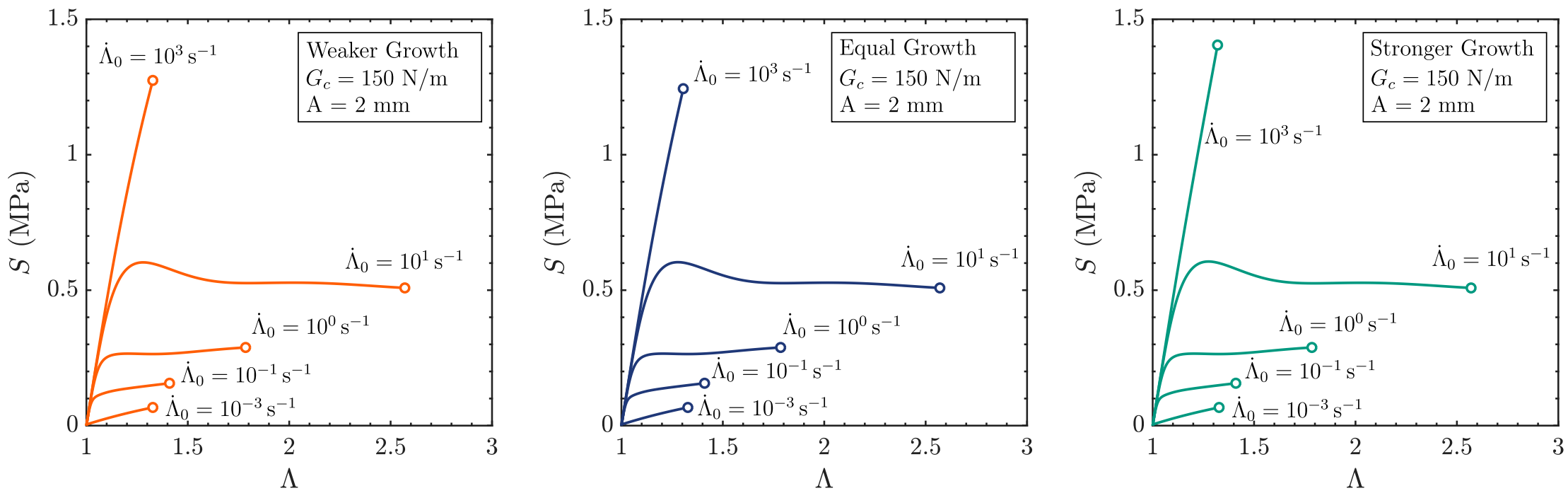

**Figure A.2** Force-deformation response of specimens with pre-existing cracks of length $A = 2$ mm stretched at various constant global stretch rates $\dot{\Lambda}_0$. The results pertain to elastomers with shear-thinning viscosity (49b) and non-equilibrium elasticity of weaker (47a) , equal (47b) , and stronger (47c) growth conditions than their equilibrium elasticity.

results pertain to specimens with initial crack size $A = 2$ mm, elastomers with non-equilibrium elasticity of weaker, equal, and stronger growth conditions (47) , with $\nu = \mu^{\mathrm{NEq}} = 20\mu^{\mathrm{Eq}} = 2$ MPa, than their equilibrium elasticity, and shear-thinning viscosity (49b) with $\eta_0 = \nu\tau_0 = 2$ MPa s. All the results are plotted up to the critical global stretch $\Lambda_c$ at which the Griffith criticality condition (39) is satisfied (empty circle) for the representative case when the critical energy release rate is $G_c = 150$ N/m.

Figure A.3 presents results—that complement those presented in Figure 9 in the main body of the text—for the global stress $S = P/(HB)$ as a function of the applied global stretch $\Lambda = l/L$ for specimens stretched at the constant global stretch rates $\dot{\Lambda}_0 = 10^{-3}, 10^{-1}, 10^{0}, 10^{1}, 10^{3}$ s$^{-1}$. The results pertain to specimens with initial crack size $A = 2$ mm, elastomers

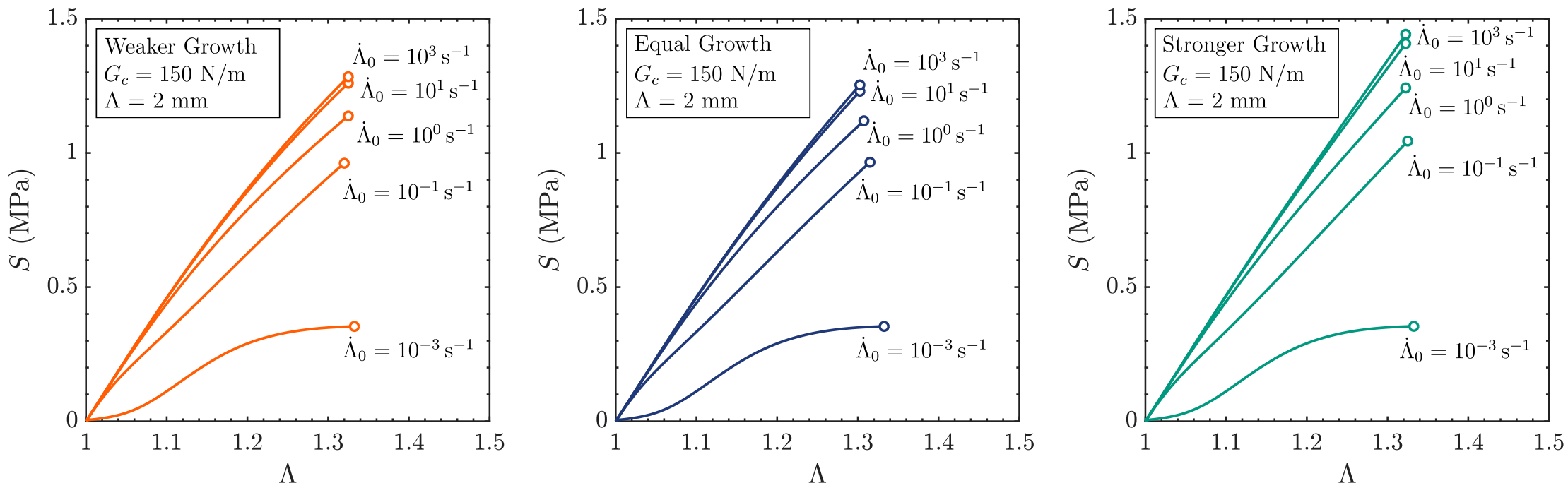

**Figure A.3** Force-deformation response of specimens with pre-existing cracks of length $A = 2$ mm stretched at various constant global stretch rates $\dot{\Lambda}_0$. The results pertain to elastomers with deformation-dependent viscosity (49c) and non-equilibrium elasticity of weaker (47a) , equal (47b) , and stronger (47c) growth conditions than their equilibrium elasticity.

with non-equilibrium elasticity of weaker, equal, and stronger growth conditions (47) , with $\nu = \mu^{\mathrm{NEq}} = 20\mu^{\mathrm{Eq}} = 2$ MPa, than their equilibrium elasticity, and deformation-dependent viscosity (49c) with $\eta_0 = \nu\tau_0 = 2$ MPa s. All the results are plotted up to the critical global stretch $\Lambda_c$ at which the Griffith criticality condition (39) is satisfied (empty circle) for the representative case when the critical energy release rate is $G_c = 150$ N/m.

**Authors' contributions** Farhad Kamarei: Writing – review & editing, Visualization, Software, Methodology; Fabio Sozio: Writing – review & editing, Visualization, Software, Methodology; Oscar Lopez-Pamies: Writing – original draft, Writing – review & editing, Visualization, Supervision, Funding acquisition, Methodology, Conceptualization

**Supplementary Material** Scripts and datasets to reproduce the proposed investigation are available at the permalink 10.5281/zenodo.20426210.

**Funding** This work was supported by the National Science Foundation (USA) through the Grant DMS–2308169. This support is gratefully acknowledged.

**Competing interests** The authors declare that they have no competing interests.

**Journal's Note** JTCAM remains neutral with regard to the content of the publication and institutional affiliations.